\documentclass[%
 reprint,
floatfix,
superscriptaddress]{revtex4-1}
\usepackage[english]{babel}
\usepackage{siunitx}

\usepackage[caption=false]{subfig}
\usepackage{floatrow}
\usepackage{amsfonts}
\usepackage{microtype}
\usepackage{amsmath}
\usepackage{mathtools}
\usepackage{tabularx}
\usepackage{appendix}
\usepackage{chngcntr}
\usepackage{graphicx}
\usepackage{dcolumn}
\usepackage{bm}
\usepackage[colorlinks=true, allcolors=blue, pdfencoding=auto]{hyperref}

\usepackage[Symbolsmallscale]{upgreek}

\addto\extrasenglish{%
}

\addto\extrasenglish{%
}

\addto\extrasenglish{%
}

\addto\extrasenglish{%
}

\addto\extrasenglish{%
}
\addto\extrasenglish{%
}

\usepackage[capitalize,nameinlink]{cleveref}
\creflabelformat{equation}{#2#1#3}
\crefname{section}{Sec.}{Secs.}
\begin{document}


\title{Modeling emission of acoustic energy during bubble expansion in PICO bubble chambers}

\author{Tetiana Kozynets}%
 \email{kozynets@ualberta.ca}
\author{Scott Fallows}
\author{Carsten B. Krauss}
\affiliation{ 
Department of Physics, University of Alberta, Edmonton, T6G 2E1, Canada
}

\date{July 25, 2019}

\begin{abstract}
The PICO experiment uses bubble chambers filled with superheated C$_3$F$_8$ for spin-dependent WIMP dark matter searches. One of the main sources of background in these detectors is alpha particles from decays of environmental $^{222}\mathrm{Rn}$, which nucleate bubbles that are visually indistinguishable from WIMP candidate events. Alpha-induced bubbles can be discriminated acoustically, because the signal from alpha events is consistently larger in magnitude than that from nuclear recoil/WIMP-like events. By studying the dynamics of bubbles nucleated by these two types of ionizing radiation from the first stages of their growth, we present a physical model for the acoustic discrimination for the first time. The distribution of acoustic energies that we generate for a simulated sample of bubble nucleations by alpha particles and nuclear recoils is compared directly to the experimental data.

\end{abstract}

\maketitle

\section{Introduction}\label{sec:intro}

Bubble chambers played an important role in experimental advances of particle physics in the 1950s and 1960s \cite{glaser_characteristics_1955}. Initially, they were operated at very high superheats to allow reliable bubble nucleation by minimally ionizing particles. Over the last decade bubble chambers have made a resurgence in dark matter searches, where the chambers are operated at more moderate superheats that explicitly prevent nucleation from minimally ionizing particles. This way, the chambers are only sensitive to nuclear recoils and other highly ionizing particle interactions. 

The PICO experiment searches for WIMP dark matter candidates using bubble chamber technology  \cite{amole_dark_2017, amole_dark_2019, mitra_pico-60:_2018}. To calibrate the detector, neutron radiation from a source of known activity, typically $^{252}\mathrm{Cf}$ or $^{241}\mathrm{Am}/^{9}\mathrm{Be}$, is used. Neutron source calibrations allow the observation of nuclear recoil events in the range of keV recoil energies, creating an acoustic signature that is identical to the expected recoil signal from dark matter interactions. 

Alpha radiation from decays of environmental $^{222}\mathrm{Rn}$ is one of the dominant sources of background. Visually, bubbles nucleated by $\alpha$ particles are indistinguishable from nuclear recoil/WIMP-like events. It was discovered in the PICASSO experiment that the acoustic signature of a nucleation from an $\alpha$ event is more powerful than the acoustic signal of a nuclear recoil event \cite{aubin_discrimination_2008}. So far this has been explained with a model that lacked specific predictability.  To reveal the origin of the observed differences in acoustic signal magnitudes, we connect the initial thermodynamic conditions of bubble nucleation by $\alpha$ particles and nuclear recoils with the specifics of bubble evolution. We approach this problem with a molecular dynamics simulation, which lets us vary the energy deposited by an incident ion and the shape of the initially vaporized region to reproduce $\alpha$-like and nuclear recoil-like scenarios. We further study the influence of these variables on the dynamics of a nucleated nonspherical bubble and analytically link the first expansion stages to the well-described spherical bubble growth \cite{plesset_growth_1954, rayleigh_pressure_1917}. Finally, we reconstruct the acoustic pressure signal from the spherical bubble growth histories for both $\alpha$-induced and nuclear recoil-induced bubbles. The results of the developed model are compared to recent acoustic data from the PICO-60 detector \cite{amole_dark_2017,amole_dark_2019}, which used superheated C$_3$F$_8$ as the target fluid.

\section{Inputs to the acoustic emission model}\label{sec:methods}

\subsection{Molecular dynamics (MD) simulations}\label{sec:simsetup}
To model the nucleation and subsequent expansion of bubbles in superheated C$_3$F$_8$, we make use of the Large-Scale Atomic/Molecular Massively Parallel Simulator (LAMMPS) \cite{plimpton_fast_1995}. Inside this package, we represent C$_3$F$_8$ as a system of atoms interacting according to the Lennard-Jones (LJ) potential with characteristic energy $\epsilon$ and characteristic distance $\sigma$ (see \cref{appendix:appendix_lj_parameters} for details). This method neglects interatomic interactions within the molecules themselves, following the approach of  \cite{denzel_molecular_2016} to model C$_2$ClF$_5$ used in the SIMPLE experiment \cite{the_simple_collaboration_final_2012}. The Lennard-Jones potential description of any fluid is complete when the $\epsilon$ and $\sigma$ parameters are specified. Setting $\epsilon = \SI{0.0318}{eV}$ makes the critical temperature of an LJ fluid match that of C$_3$F$_8$. We also aim to simulate the specific conditions of the PICO-60 run at temperature $T_0 \simeq \SI{14}{\celsius}$ and liquid pressure $P_{\mathrm{l}} \simeq \SI{207}{kPa}\,(\SI{30}{psia})$. This corresponds to the C$_3$F$_8$ liquid density of $\rho_{\mathrm{l}} = \SI{1379}{kg}\,\mathrm{m^{-3}}$, which we can reproduce with $\sigma = \SI{0.533}{nm}$. At \SI{14}{\celsius}, such a choice results in a $16\%$ discrepancy between the REFPROP \cite{LEMMON-RP91} value of equilibrium pressure at saturation and the Lennard-Jones value measured from our MD simulations; $33\%$ between REFPROP-extracted and simulated vapor densities; and $25\%$ between the surface tension values. \cref{appendix:appendix_lj_parameters} summarizes the steps we took to arrive at these results.

To bring the LJ fluid approximating C$_3$F$_8$ to the specified superheated state, we follow the procedure described in \cite{denzel_molecular_2016} and outlined in \cref{appendix:heat_spike_simulation}. As in \cite{denzel_molecular_2016}, we proceed with simulating the heat spike by depositing energy $E_{\mathrm{dep}}$ within a narrow and long cylindrical region in the metastable liquid, mimicking the track of an ionizing particle. Practically, this is done by increasing the velocities of the atoms falling within the cylindrical region to the corresponding temperature. According to the Seitz model \cite{seitz_theory_1958}, our chosen thermodynamic conditions require a minimum of $\mathcal{T}_{\mathrm{min}} = 3.3\,\mathrm{keV}$ to be deposited within the critical bubble radius $R_{\mathrm{c}}$ for the nucleated bubble to continue growing stably. To imitate nuclear recoils, we perform several simulations with the cylinder length $l_{\mathrm{cyl}}$ equal to integer multiples of $R_{\mathrm{c}}$ ($40.58 \sigma \approx$ \SI{21.63}{nm} at our conditions) and energy deposits close to $\mathcal{T}_{\mathrm{min}}$. The radius of the cylindrical energy deposition region is fixed at $r_{\mathrm{cyl}} = 2\sigma \approx 1.066\,\si{nm}$. To simulate $\alpha$-like bubble nucleations, we deposit energies of order \SI{10}{keV} along \SI{}{\micro\meter}-long cylindrical tracks, with $r_{\mathrm{cyl}} = 3\sigma \approx 1.599\,\si{nm}$. We justify these geometry choices in \cref{appendix:heat_spike_simulation}.

The simulated bubbles are allowed to grow under \textit{NPT} (constant particle number, pressure, and temperature) constraints through several major stages. These include rapid initial bubble expansion, followed by slight decrease in the bubble volume over a short time interval, and subsequent transition to the inertial growth phase. In this phase, the effective bubble radius $R_{\mathrm{eff}}$ increases nearly linearly with time, being controlled by the forces of liquid inertia. When the bubble, which grew out of a long cylindrical vapor region, becomes spherical, we may predict its further behavior by solving the differential equation describing spherical bubble growth \cite{plesset_growth_1954,rayleigh_pressure_1917}. This technique allows us to simulate only the first few tens of nanoseconds of bubble expansion in LAMMPS, feeding the MD-based $R(t)$ history into the familiar expressions for $R$ at later times. In \autoref{sec:alpha_bubbles}, we show that $\alpha$-induced bubbles transition from nonspherical to spherical in a similar fashion, meaning that the same approach is applicable to predicting their growth.

\subsection{Post-simulation processing: Bubble surface tracking}\label{sec:bubble_identification}

\begin{figure}[h!]
  \includegraphics[scale=0.28]{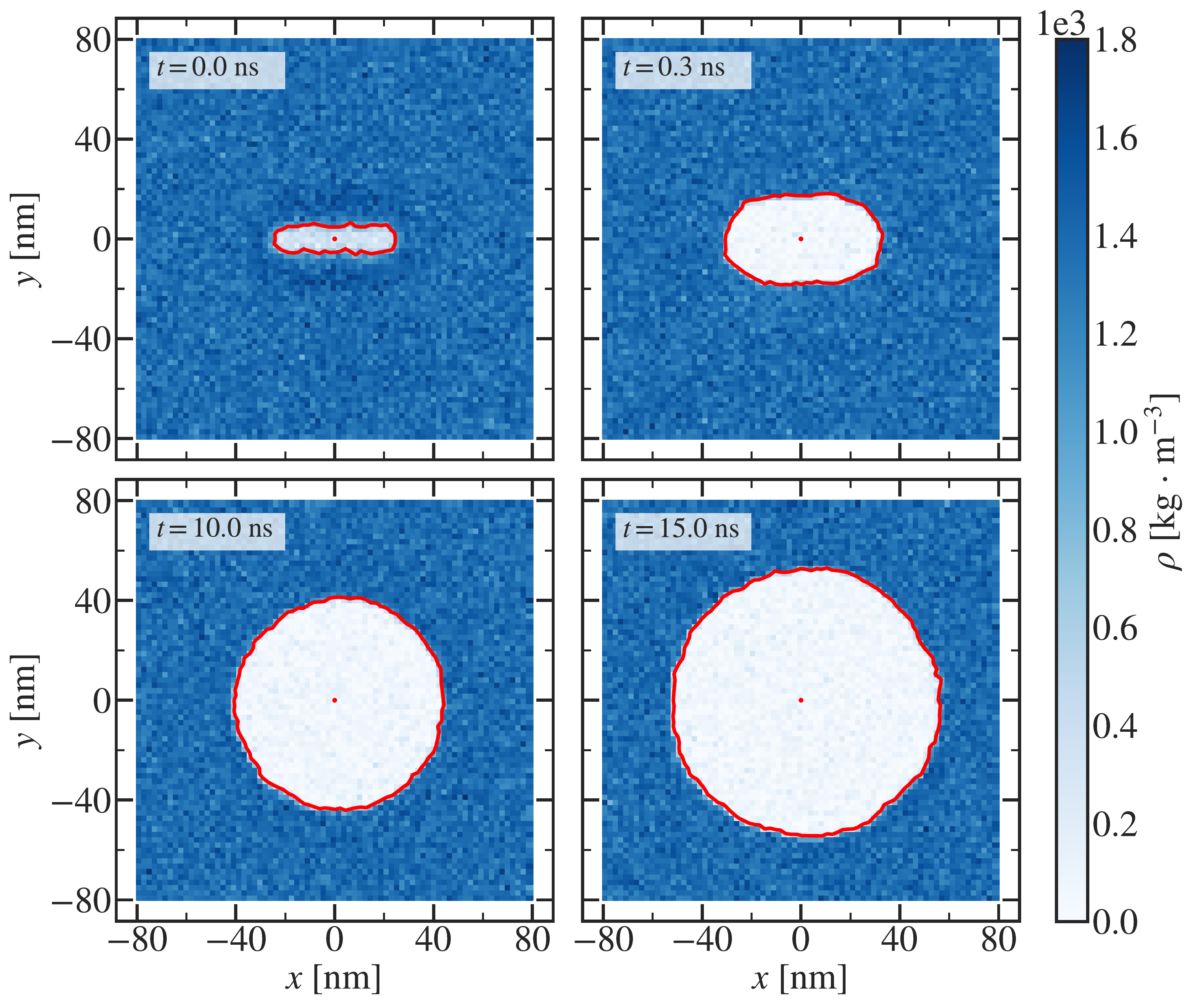}
  \centering
  \caption{Example time evolution of a simulated bubble (this case: $E_{\mathrm{dep}}$ = \SI{3.75}{keV} deposited along $l_{\mathrm{cyl}} = 2R_{\mathrm{c}}$, parallel to the $x$-axis). The plots show positions of the grid cells in the $XY$ plane, color-coded by the time-averaged fluid densities in the cells. The solid red line is the longest contour extracted with the marching squares algorithm \cite{lorensen_marching_1987,walt_scikit-image:_2014} from the density data, defining the bubble surface boundary in each case.}
  \label{fig:bubble_surface_tracking}
\end{figure}

From each molecular dynamics simulation, we save 2D distributions of liquid densities in the \textit{XY} plane running through the middle of the simulation volume (see \cref{appendix:surface_tracking_appendix} for details). With energy deposited along the $x$-axis, the \textit{XY} plane fully captures the 3D bubble growth, given that the $z$-axis is physically no different from the $y$-axis. These density distributions are then used as an input to the marching squares algorithm \cite{lorensen_marching_1987, walt_scikit-image:_2014}, which allows us to extract the contours of constant density values. As per \cref{appendix:surface_tracking_appendix}, the longest contour of $\rho_{\mathrm{l}} \approx \SI{830}{kg}\,\mathrm{m^{-3}}$ accurately reproduces the liquid-vapor boundary at each $t$. The $(x,y)$ coordinates of this boundary let us evaluate the bubble radius as a function of polar angle $\phi \equiv \tan^{-1}(y/x)$. In the present work, we take advantage of this to track the eccentricity of bubbles and thereby define the time elapsed after the heat spike when each bubble can be considered nearly spherical. As discussed in \cref{sec:mikic_bubble_evolution}, specifying this time and the corresponding value of the bubble radius makes for an initial condition sufficient to predict the subsequent bubble evolution. In \autoref{fig:bubble_surface_tracking}, we give an example of bubble boundary extraction results at several representative times after the heat spike.

\subsection{Predicting evolution of a spherical bubble in the linear and thermal growth phases}\label{sec:mikic_bubble_evolution}

Molecular dynamics simulations, even for liquid volumes as small as those considered so far, are quite computationally expensive: for instance, $\sim$\SI{1}{ns} of a typical bubble evolution in an \textit{NPT} ensemble takes about 720 CPU-hours to run in LAMMPS. At the same time, radiation of acoustic energy as the result of bubble growth spans much longer time scales. These are roughly defined by the time of transition from the inertial bubble growth phase to the thermal growth phase, which features much slower expansion and consequently smaller amplitudes of the emitted sound waves \cite{landau_fluid_2013}. For a bubble growing in superheated C$_3$F$_8$, the characteristic inertial growth phase duration is of order $\SI{100}{\micro\second}$ \cite{mitra_pico-60:_2018}. This makes simulating the whole acoustic emission practically impossible with molecular dynamics methods alone. Such a complication calls for a way to fully predict bubble evolution based on a short history of its growth. The method has to be further generalized to the case of an arbitrary energy deposition, letting us perform MD simulations for only a few different $E_{\mathrm{dep}}$ values and extrapolate the results to other energies. 

\begin{figure}
    \subfloat[Bubble radius as a function of time elapsed after the simulated energy deposition of \SI{5}{keV} along $l_{\mathrm{cyl}} = 2R_{\mathrm{c}} \approx \SI{43.3}{nm}$, extracted at different angles $\phi$ with the energy deposition direction (see \autoref{sec:bubble_identification}).]{\label{radius_vs_time}\includegraphics[scale = 0.3]{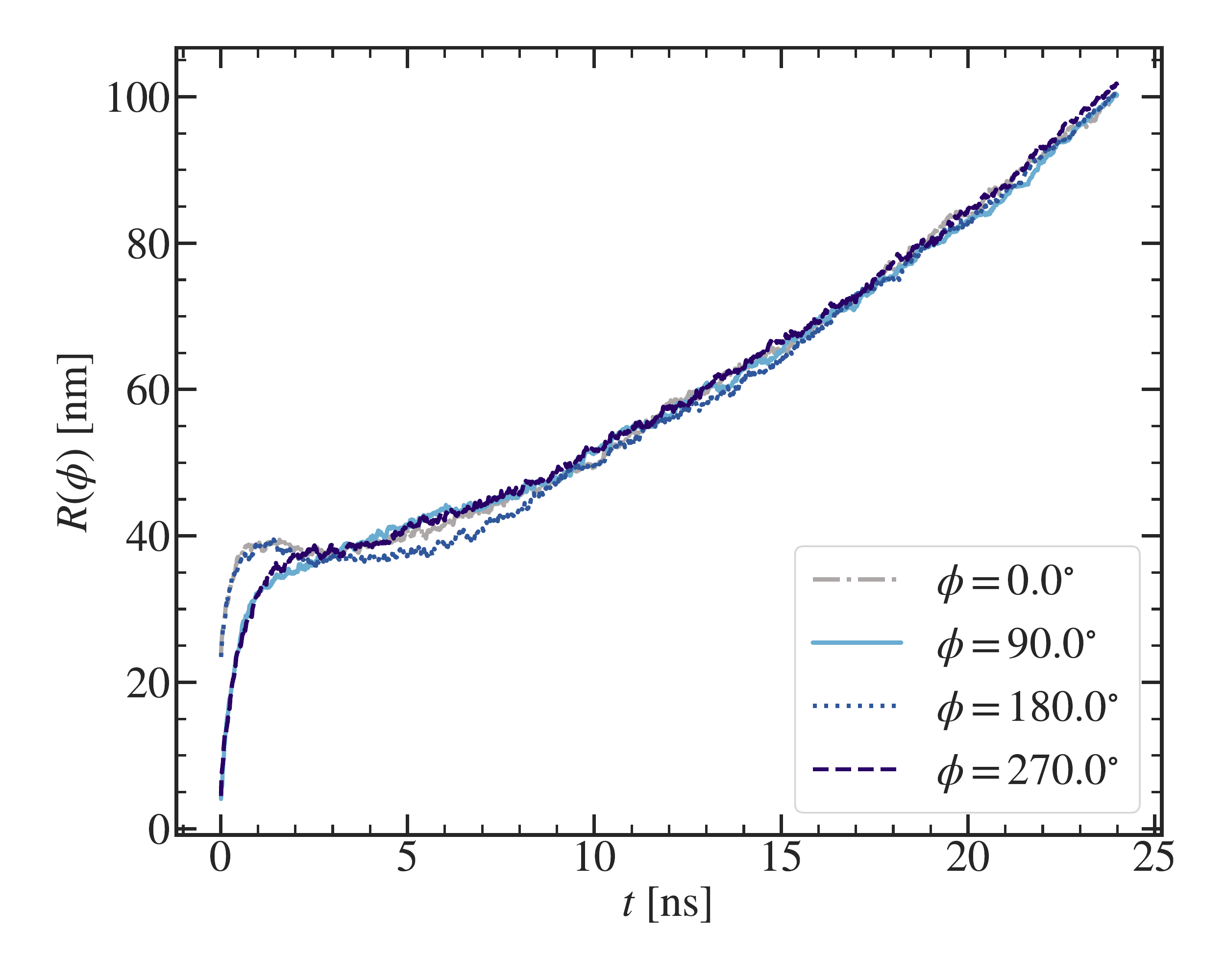}} \\
    \subfloat[Bubble eccentricity \eqref{eq:eccentricity_vs_time} derived from the $R(t)$ data in \autoref{radius_vs_time}. The first minimum (the circled red cross) is taken as the definition of time $t_{\mathrm{s}}$ after which the bubble can be treated as spherical.]{\label{fig:eccentricity_vs_time}\includegraphics[scale = 0.3]{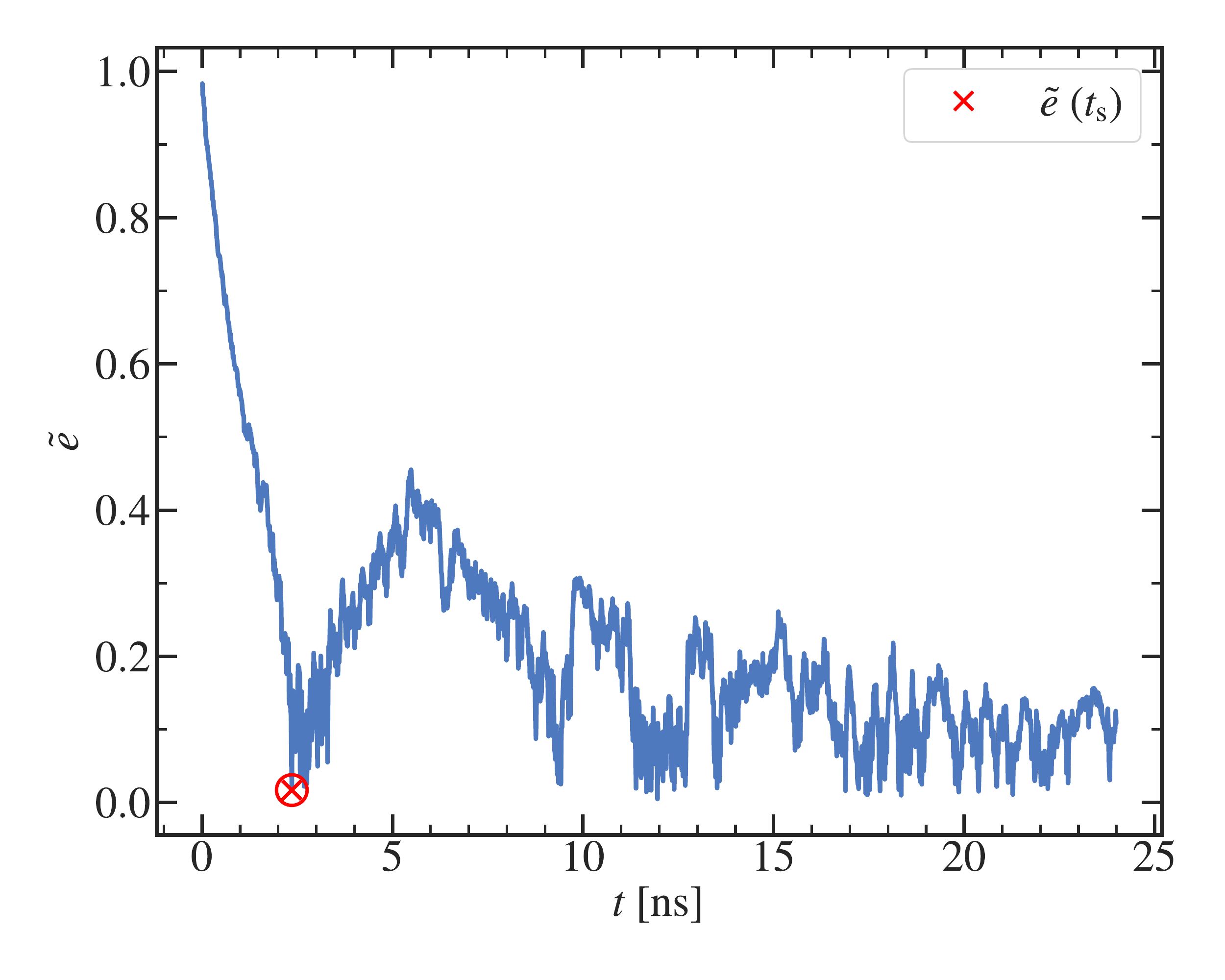}}
    \caption{The strategy for defining the range of applicability of the spherical bubble growth equation \eqref{eq:full_thermal_DE} to the evolution of a bubble simulated in the superheated C$_3$F$_8$-like LJ fluid.}
    \label{fig:fitting_strategy}
\end{figure}

To proceed, we need to make a connection between bubble nucleation conditions, growth of the resulting highly nonspherical bubbles, and well-described spherical bubble dynamics. By combining the Rayleigh-Plesset equation \cite{rayleigh_pressure_1917} for the inertial growth phase and the Plesset-Zwick equation \cite{plesset_growth_1954} for the thermal growth phase as in \cite{mikic_bubble_1970}, we arrive at the bubble growth rate
\begin{equation}
\begin{split}
\frac{\mathrm{d}R}{\mathrm{d}t} = -\left[\frac{A^2 \sqrt{t - t_{\mathrm{s}}}}{B} + \frac{2 \nu_{\mathrm{l}}}{R}\right] 
\\ &\hspace*{-95pt} + \sqrt{A^2 - \frac{2 \gamma}{\rho_{\mathrm{l}} R} + \left(\frac{2 \nu_{\mathrm{l}}}{R} + \frac{A^2 \sqrt{t - t_{\mathrm{s}}}}{B} \right)^2}.
    \label{eq:full_thermal_DE}
\end{split}
\end{equation} 
In \autoref{eq:full_thermal_DE}, the constant $A$ is a characteristic speed of expansion in the inertial phase. It is given by

\begin{equation}
    \label{eq:definition_of_A}
    A = \sqrt{\frac{2}{3} \frac{h \rho_{\mathrm{v}} \Delta T}{\rho_{\mathrm{l}} T_{\mathrm{sat}}}},
\end{equation}
where $h$ is the latent heat of vaporization, $\Delta T$ is the liquid superheat above the saturation temperature $T_{\mathrm{sat}}$ for a given liquid density $\rho_{\mathrm{l}}$, and $\rho_{\mathrm{v}}$ is the vapor density inside the bubble. The constant $B$ in \eqref{eq:full_thermal_DE} may be expressed as 
\begin{equation}
    \label{eq:definition_of_B}
    B = \sqrt{\frac{12}{\pi a_{\mathrm{l}}}} \textit{Ja},
\end{equation}
where $a_{\mathrm{l}}$ is the thermal diffusivity of the liquid and $\textit{Ja}$ is its Jakob number ($\textit{Ja} = \frac{\Delta T c_{\mathrm{l}} \rho_{\mathrm{l}}}{h \rho_{\mathrm{v}}}$, with  $c_{\mathrm{l}}$ being the liquid heat capacity.) The other quantities appearing in \autoref{eq:full_thermal_DE} are the liquid viscosity $\nu_{\mathrm{l}}$ and the surface tension $\gamma$, whose contributions are non-negligible for the small bubble sizes simulated in LAMMPS. In \autoref{tab:bubble_growth_parameters}, we provide the REFPROP values of these properties for C$_3$F$_8$ at \SI{14}{\celsius} and \SI{207}{kPa}\,(\SI{30}{psia}), as well as the $A$ and $B$ constants evaluated according to \cref{eq:definition_of_A,eq:definition_of_B}. For our C$_3$F$_8$-like LJ fluid, we may also find  $A, B, \gamma$, and $\nu_{\mathrm{l}}$  by fitting that part of the simulation-extracted $R(t)$ data where the bubble is spherical.
To determine when this happens, we examine the time evolution of eccentricity, defined as
\begin{equation}
    \label{eq:eccentricity_vs_time}
    \tilde{e} = \sqrt{\left|1 - \frac{R_{90}^2}{R_{0}^2}\right|},
\end{equation}
with $R_0$ being the bubble radius in the direction of energy deposition ($\phi = 0^{\circ}$) and $R_{90}$ being that in the orthogonal direction ($\phi$ = 90$^{\circ}$). The bubble eccentricity as a function of time corresponding to $R(t)$ from \autoref{radius_vs_time} is shown in \autoref{fig:eccentricity_vs_time}. We choose to use the first minimum in bubble eccentricity as an approximation for the time $t_{\mathrm{s}}$ required for the bubble to become spherical. Each $t_{\mathrm{s}}$ corresponds to a certain bubble radius value $R_{\mathrm{s}}$. $R(t_{\mathrm{s}}) = R_{\mathrm{s}}$ serves as the initial condition for \autoref{eq:full_thermal_DE}, which we fit simultaneously to the four simulated data sets as described in \cref{appendix:sphericity_radii_and_time_relations}. With parameter $A$ fixed at its value predicted for C$_3$F$_8$ with \autoref{eq:definition_of_A}, we find a satisfactory agreement between the best-fitting $B$ and $\gamma$ and their C$_3$F$_8$ values (see \autoref{tab:bubble_growth_parameters}). The viscosity of C$_3$F$_8$ differs from the value we extract for our LJ fluid by more than two orders of magnitude. This is a natural result since the Lennard-Jones potential completely neglects the intramolecular interactions within C$_3$F$_8$ and cannot reproduce all of its thermodynamic properties at once. We choose to proceed with the simulation-extracted rather than REFPROP parameter values in order to develop a self-consistent description of bubble expansion in the C$_3$F$_8$-like Lennard-Jones fluid. We stress that the bubble dynamics and acoustic emission model is therefore constructed for such an LJ fluid and not C$_3$F$_8$ itself. However, we find that replacing the best-fitting LJ viscosity value with the REFPROP value for C$_3$F$_8$ would eventually result in $\mathcal{O}(1)$ scaling of the integrated acoustic power in our frequency range of interest (\SIrange[range-phrase=--,range-units=single]{1}{300}{kHz}; see \autoref{sec:power_extraction}) for both nuclear recoil- and $\alpha$-like bubble nucleations. Since both types of simulated particle interaction are affected in the same way, the relative behavior between the two cases remains well described.

\begin{table}
\caption{Select thermodynamic properties of C$_3$F$_8$ at \SI{14}{\celsius} and \SI{207}{kPa}\,(\SI{30}{psia}) extracted from REFPROP \cite{LEMMON-RP91}. When appropriate, we also list the respective values found for the C$_3$F$_8$-like LJ fluid and use a dagger$^{\dagger}$ to denote these cases.}
\label{tab:bubble_growth_parameters}
\begin{tabular*}{\textwidth}{l @{\extracolsep{\fill}} ll}
Property & Value & Unit \\[3pt]\hline\\[-1.5ex]
$h$: enthalpy of vaporization & $82.58$ & $\mathrm{{kJ}\,{kg}^{-1}}$ \\[3.5pt]
$c_{\mathrm{l}}$: liquid heat capacity   & $1130.5$ & $\mathrm{{J}\,{kg^{-1}\,K^{-1}}}$ \\[3.5pt] 
$a_{\mathrm{l}}$: liquid thermal diffusivity & $0.030$ & $\mathrm{{mm^2}\,{s}^{-1}}$ \\ [3.5pt]
$\rho_{\mathrm{v}}$: vapor density & $59.4$ & $\mathrm{{kg}\,{m^{-3}}}$\\[3.5pt]
$\rho_{\mathrm{l}}$: liquid density & $1379.0$ & $\mathrm{{kg}\,{m^{-3}}}$\\[3.5pt]
$T_{\mathrm{sat}}$: saturation temperature & 253.57 & K \\[3.5pt]
 $A$ (\autoref{eq:definition_of_A}) & $17.7$ & $\mathrm{{m}\,{s}^{-1}}$ \\[3.5pt]
 $B$ (\autoref{eq:definition_of_B}) & $0.0036$ & $\mathrm{{m}\,{s^{-1/2}}}$ \\[3.5pt]
  & $0.0086 \pm 0.0002^{\dagger}$ & $\mathrm{{m}\,{s^{-1/2}}}$ \\[3.5pt]
$\nu_{\mathrm{l}}$: kinematic viscosity & $0.141$ & $\mathrm{{mm^2}\,{s}^{-1}}$ \\[3.5pt]
 & $18.3\pm0.4^{\dagger}$ & $\mathrm{{mm^2}\,{s}^{-1}}$ \\[3.5pt]
$\gamma$: surface tension & $4.9$ & $\mathrm{{mN}\,{m}^{-1}}$ \\[3.5pt]
 & $3.667 \pm 0.007^{\dagger}$ & $\mathrm{{mN}\,{m}^{-1}}$ \\[3.5pt]

\end{tabular*}
\end{table}

\begin{figure}
  \includegraphics[scale=0.33]{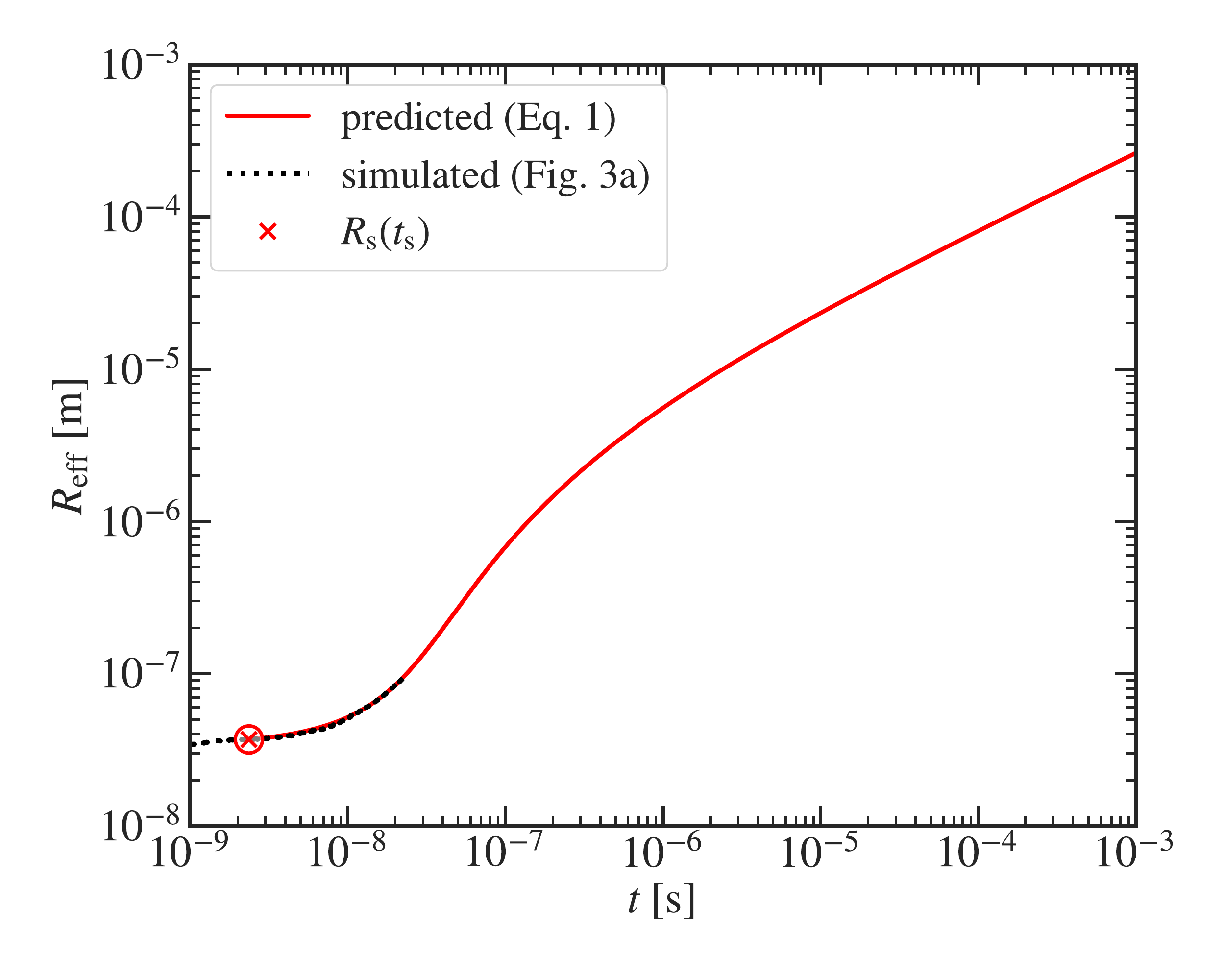}
  \centering
  \caption{Time evolution of effective bubble radius $R_{\mathrm{eff}}$ for a bubble nucleated from the energy deposition of $E_{\mathrm{dep}} = 5\,\mathrm{keV}$ over $l_{\mathrm{cyl}} = 2R_{\mathrm{cyl}} \approx 43.3\,\mathrm{nm}$. We show the bubble growth history extracted from LAMMPS (dotted black line) and that predicted with \autoref{eq:full_thermal_DE} (solid red line), with the red cross denoting the moment when the bubble became spherical.}
  \label{fig:log-log_r_vs_time}
\end{figure}

With all parameters in \autoref{eq:full_thermal_DE} determined for our LJ fluid, we can predict the evolution of bubbles nucleated within it for all $t \geq t_{\mathrm{s}}$. In \autoref{fig:log-log_r_vs_time}, we give an example of such a prediction for a bubble nucleated from $E_{\mathrm{dep}} = \SI{5}{keV}$ deposited along $l_{\mathrm{cyl}} = 2R_{\mathrm{c}}$. However, it remains to be shown how exactly the initial condition $R(t_{\mathrm{s}}) = R_{\mathrm{s}}$ varies with $E_{\mathrm{dep}}$ and $l_{\mathrm{cyl}}$. As per \cref{appendix:sphericity_radii_and_time_relations}, we find 

\begin{equation}
    \label{eq:definition_of_sphericity_time}
    t_{\mathrm{s}}~[\mathrm{ns}] \approx (0.082 \pm 0.006)l_{\mathrm{cyl}} ~[\mathrm{nm}]
\end{equation}
\noindent and
\begin{equation}
\begin{split}
R_{\mathrm{s}}~[\mathrm{nm}] \approx (0.38 \pm 0.03)\, l_{\mathrm{cyl}}\,[\mathrm{nm}] 
\\ &\hspace*{-95pt} + (217 \pm 27)\, \frac{E_{\mathrm{dep}}}{l_{\mathrm{cyl}}}\,\left[\mathrm{\frac{keV}{nm}}\right].
    \label{eq:definition_of_sphericity_radius}
\end{split}
\end{equation} 

\noindent Together, \cref{eq:definition_of_sphericity_time,eq:definition_of_sphericity_radius} provide the initial condition for \autoref{eq:full_thermal_DE} given deposited energies and lengths of ion tracks, which are known for both nuclear recoils and $\alpha$ particles. 

\subsection{Predicting the growth of $\alpha$-induced bubbles}\label{sec:alpha_bubbles}

Three populations of alpha-decay events have been observed in the PICO bubble chambers:

\begin{align}
    \label{eq:decay_chains}
    ^{222}\mathrm{Rn} \to\,^{218}\mathrm{Po} + \alpha + 5.6\,\mathrm{MeV}; \nonumber \\
     ^{218}\mathrm{Po} \to \,^{214}\mathrm{Po} + \alpha + 6.1\,\mathrm{MeV}; \\
     ^{214}\mathrm{Po} \to \,^{210}\mathrm{Pb} + \alpha + 7.8\,\mathrm{MeV} \nonumber.
\end{align}

These events belong to the decay chain of environmental $^{222}\mathrm{Rn}$, which creates a background of $\alpha$ particles depositing their energy within the superheated C$_3$F$_8$. Once bubbles nucleated as the result of such an energy deposition grow to a visible size, they cannot be distinguished from neutron-induced bubbles based on visual appearance. Instead, the acoustic parameter (AP) has been introduced as a measure of total acoustic power radiated during bubble expansion \cite{amole_dark_2017}.  In the 1-300 kHz frequency range, the experimental AP provides excellent discrimination against $\alpha$-induced bubbles, which have 5-20 times larger AP values than those measured for nuclear recoil events, depending on the $\alpha$ energy. The dynamics of $\alpha$-induced bubbles has not been fully investigated before, which led to the lack of understanding of the origin of the $\alpha$ signal amplitude. Here, we will examine the dynamics of the bubbles nucleated by the $^{222}\mathrm{Rn}$ decay chain $\alpha$ particles specifically.

\begin{figure}
  \includegraphics[scale=0.3]{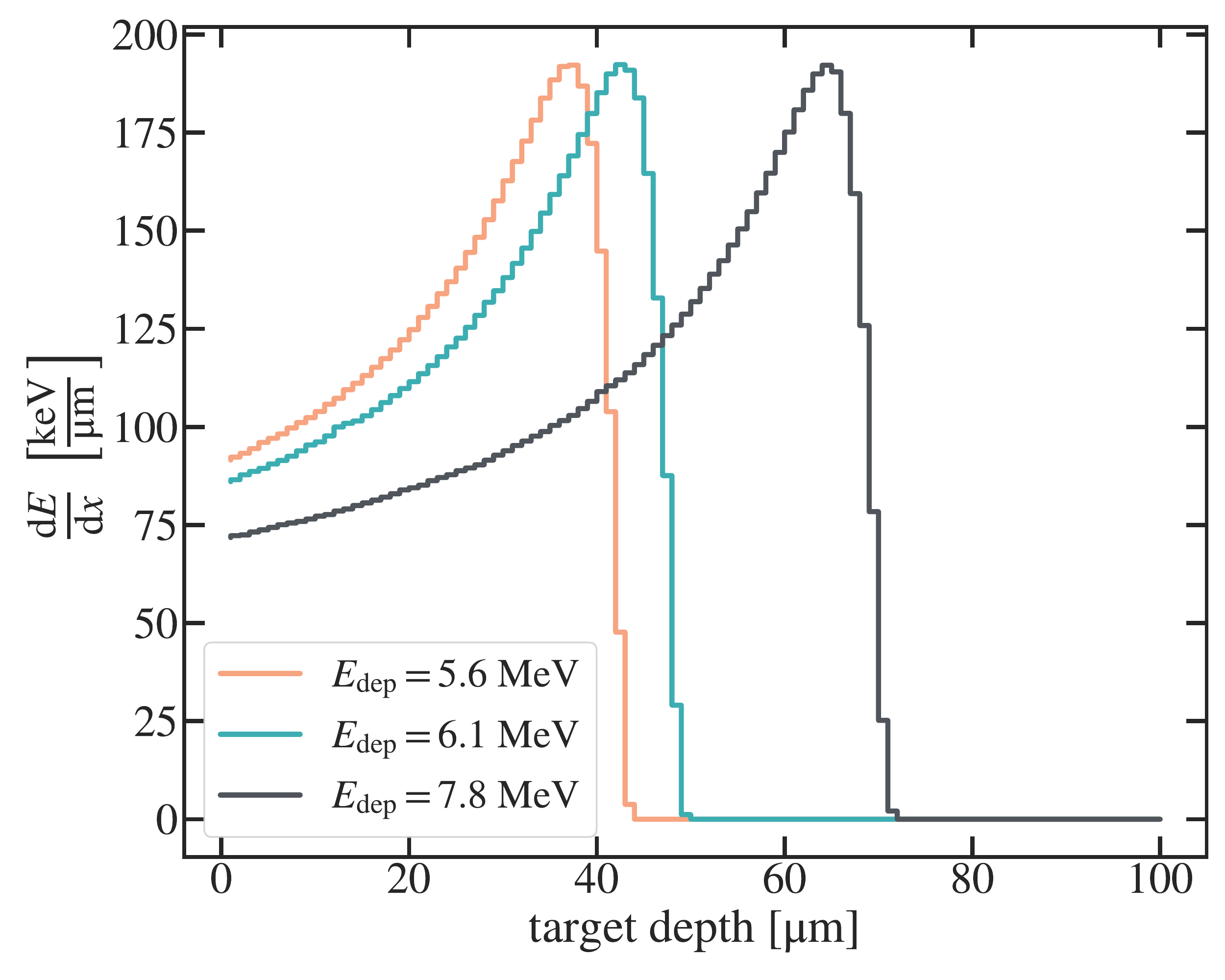}
  \centering
  \caption{Energy per unit length lost by $\alpha$ particles with initial energies of 5.6, 6.1, and 7.8\,MeV to ionizing C and F atoms in C$_3$F$_8$ ($\rho_{\mathrm{l}} = 1379\,\mathrm{kg \, m^{-3}}$), shown as a function of the depth of penetration into C$_3$F$_8$ \cite{ziegler_srim_2010}.}
  \label{fig:srim_ranges}
\end{figure}

In \autoref{fig:srim_ranges}, we plot $\frac{\mathrm{d}E}{\mathrm{d}x}$ lost to ionization as a function of $\alpha$ penetration depth in C$_3$F$_8$ of $\rho_{\mathrm{l}} = 1379\,\mathrm{kg \, m^{-3}}$. These data are extracted with the Stopping Range of Ions in Matter (SRIM) package \cite{ziegler_srim_2010} for $\alpha$ particles with energies equal to the full energies released in the decays. The recoil energies of the heavy daughter nuclei are of order \SI{100}{keV} and are deposited over tens of nanometers in close proximity to the significantly longer $\alpha$-induced vapor regions. For that reason, our model assumes that the total energy released in each decay is deposited continuously over the length of the respective $\alpha$ track. 

\begin{figure}
  \includegraphics[scale=0.25]{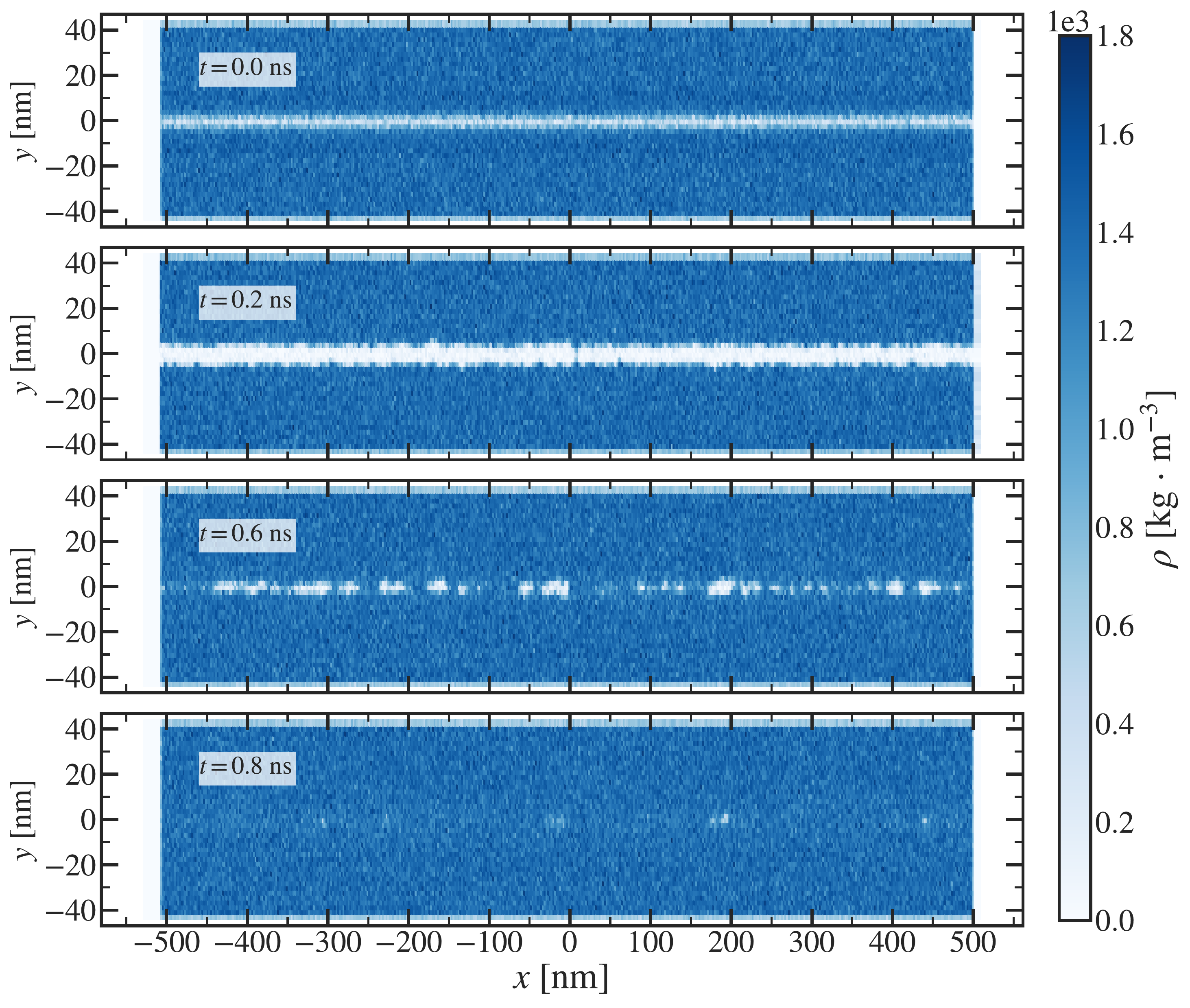}
  \centering
  \caption{Example time evolution of a $1\,\mathrm{\SI{}{\micro\meter}}$ long vapor region resulting from a \SI{10}{keV} energy deposition into this region by an $\alpha$ particle. The vapor tube splits into several protobubbles (panel 3), all of which eventually collapse due to insufficient energy deposition. Given $\alpha$ particles from \autoref{fig:srim_ranges}, only small fractions of each $\alpha$-induced bubble will collapse as shown here, with dominant portions of $\alpha$ tracks getting higher $E_{\mathrm{dep}}$ per $\mathrm{\SI{}{\micro\meter}}$ and expanding as in \autoref{fig:expanding_alpha_tubes}.}
  \label{fig:collapsing_alpha_tubes}
\end{figure}

\begin{figure}
  \includegraphics[scale=0.25]{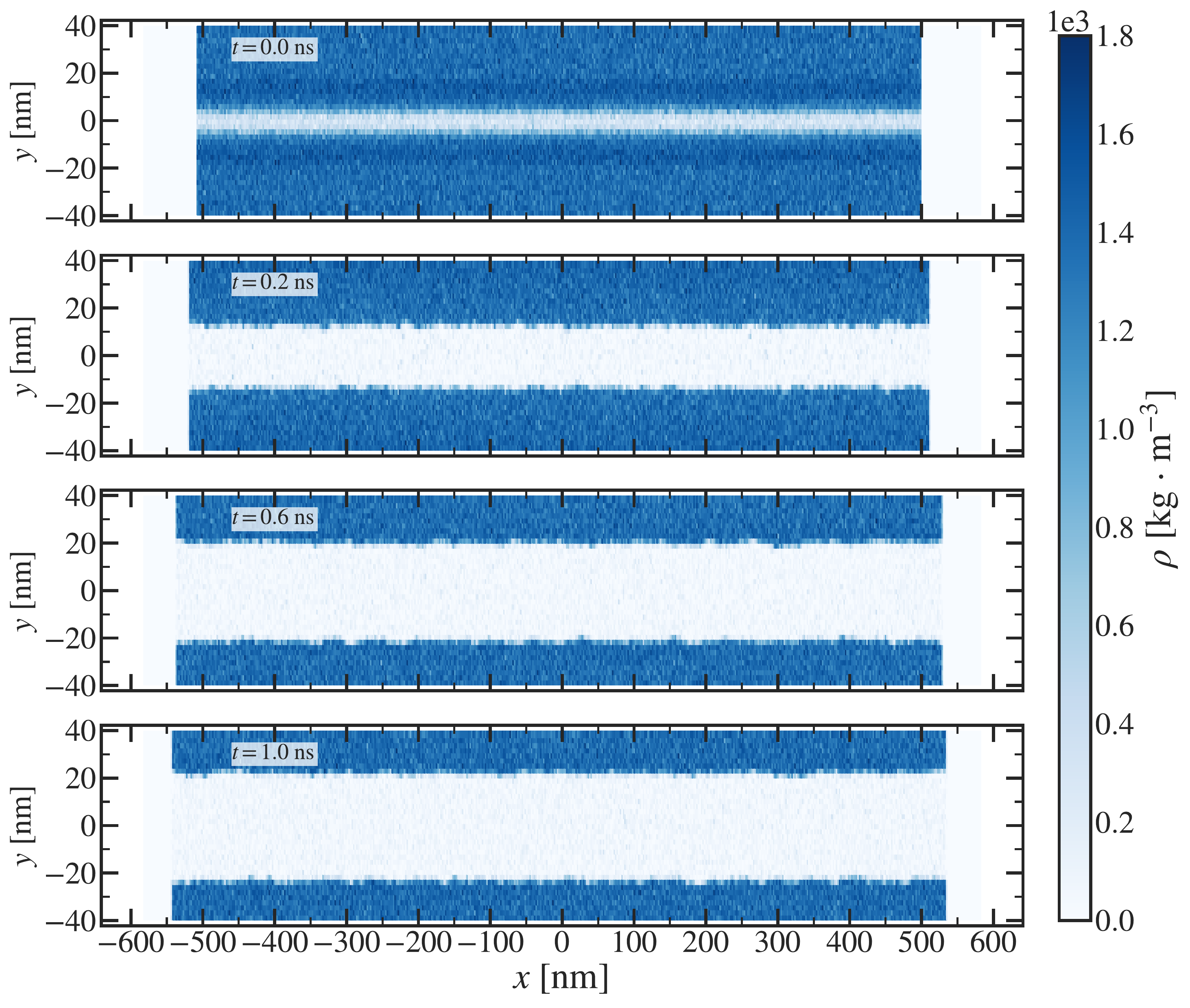}
  \centering
  \caption{Example time evolution of a $1\,\mathrm{\SI{}{\micro\meter}}$ long vapor region resulting from a \SI{50}{keV} energy deposition into this region by an alpha-particle. Unlike in \autoref{fig:collapsing_alpha_tubes}, here the vapor tube expands irreversibly, which will be the case for major fractions of $\alpha$ tracks from \autoref{fig:srim_ranges}.}
  \label{fig:expanding_alpha_tubes}
\end{figure}

\begin{figure}
  \includegraphics[scale=0.3]{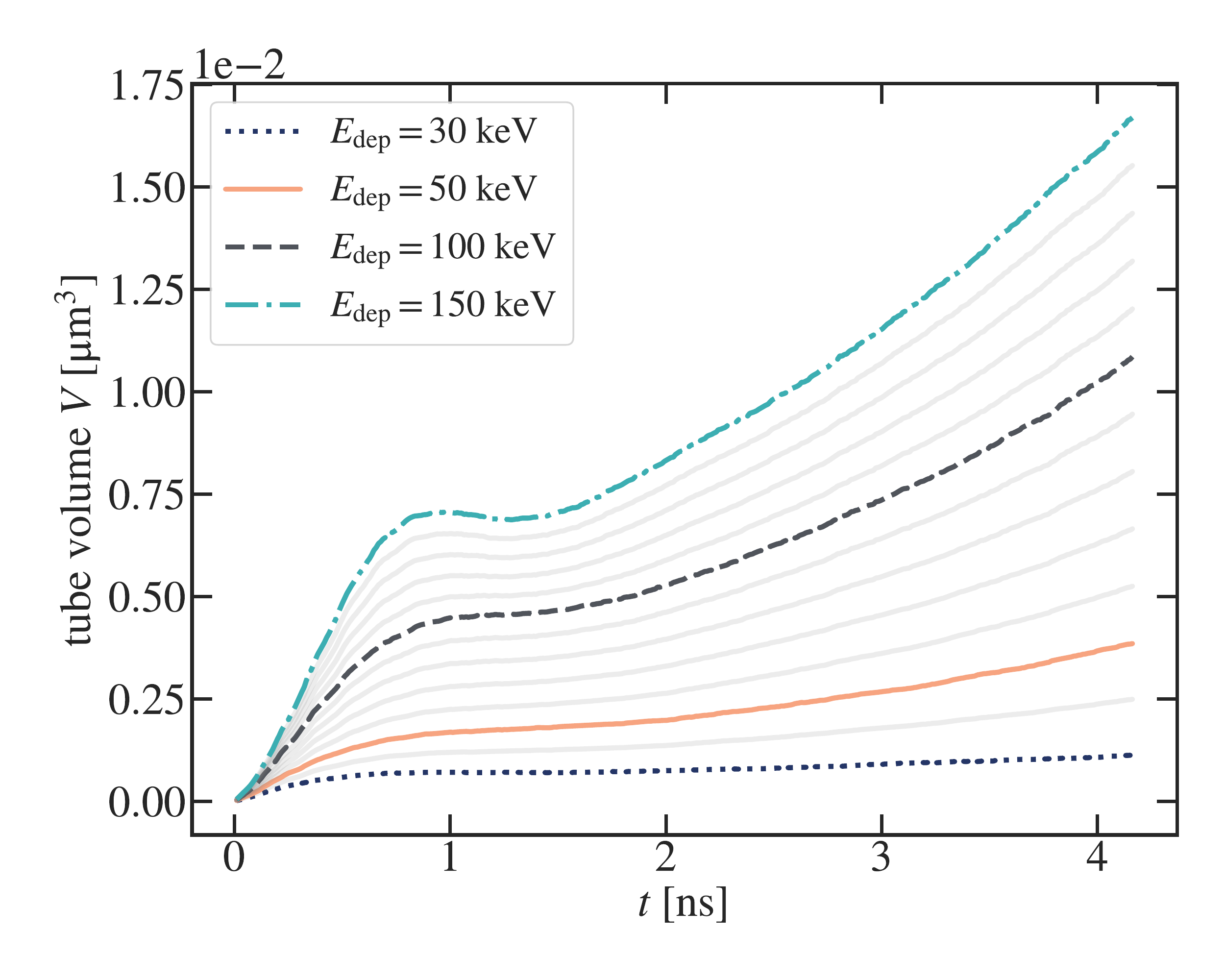}
  \centering
  \caption{Evolution of the volumes of $\mathrm{\SI{}{\micro\meter}}$-long vapor regions imitating parts of alpha-particle tracks for four attempted energy deposition values. The solid gray lines represent the $V(t)$ curves predicted for $E_{\mathrm{dep}} = N \cdot 10\,\mathrm{keV}, N \in \mathbb{N}$ by interpolation.}
  \label{fig:alpha_tubes_vol_evolution}
\end{figure}

To understand how different portions of the long $\alpha$-induced vapor tubes evolve, we simulate 10, 30, 50, 100, and \SI{150}{keV} energy depositions over \SI{1}{\micro\meter} in five separate simulations. The details of the simulation box and energy deposition region geometry are given in \cref{appendix:heat_spike_simulation}. The resulting tube of vapor is allowed to evolve in an \textit{NPT} ensemble with periodic boundary conditions on all sides. This means that each micron-long vapor chunk is evolving exactly as if it was surrounded by the same vapor chunks on both sides, imitating long alpha-particle tracks quite well. We observe that among the tested configurations, only the vapor tube produced as the result of a $10\,\mathrm{\frac{keV}{\SI{}{\micro\meter}}}$ energy deposition is separated into several protobubbles, all of which eventually collapse, as shown in \autoref{fig:collapsing_alpha_tubes}. All other tubes keep expanding irreversibly as in \autoref{fig:expanding_alpha_tubes}. In \autoref{fig:alpha_tubes_vol_evolution}, we show the evolution of the vapor tube volumes resulting from $E_{\mathrm{dep}} \geq 30\,\mathrm{keV}$. This observation implies that the bubbles induced by the three $^{222}\mathrm{Rn}$ chain $\alpha$ particles (\autoref{eq:decay_chains}) will predominantly not split along $\alpha$ ionization tracks, as the per-$\mathrm{\SI{}{\micro\meter}}$ energy depositions are smaller than \SI{30}{keV} only over the last few microns to the right of the Bragg peaks (see \autoref{fig:srim_ranges}).

This leads us to the conclusion that any bubble nucleated as the result of energy deposition by an $\alpha$ particle will evolve similarly to a bubble resulting from a nuclear recoil considered in \cref{sec:bubble_identification,sec:mikic_bubble_evolution}. If we further approximate the Bragg curves from \autoref{fig:srim_ranges} by uniform depositions of $E_{\alpha}$ (5.6, 6.1, and \SI{7.8}{MeV}) over the respective penetration depths, the evolution of the produced bubbles will be described, as in the nuclear recoil case, by \autoref{eq:full_thermal_DE}. The latter will apply at all times when an $\alpha$-induced bubble may be considered spherical, i.e. $t \geq t_{\mathrm{s}}$, where $t_{\mathrm{s}}$ is to be estimated using \autoref{eq:definition_of_sphericity_time} for $l_{\mathrm{cyl}} = 42.6, 48.3,$ and $70.0\,\mathrm{\SI{}{\micro\meter}}$ (given in the order of increasing $E_{\alpha}$). In \autoref{tab:alpha_sphericity_params}, we list the estimated times following the instant of the $\alpha$-induced heat spike at which the resulting bubbles acquire spherical shapes, as well as the corresponding bubble radii. 

\begin{table}
\caption{Estimates of the time $t_{\mathrm{s}}$ when $\alpha$-induced bubbles become spherical and their radii $R_{\mathrm{s}}$ at that moment. The values of $t_{\mathrm{s}}$ and $R_{\mathrm{s}}$ are evaluated with \cref{eq:definition_of_sphericity_time,eq:definition_of_sphericity_radius} respectively for alphas from the $^{222}\mathrm{Rn}$ decay chain \eqref{eq:decay_chains}. The reported $l_{\mathrm{cyl}}$ values correspond to the mean track ranges of $\alpha$ particles in the superheated C$_3$F$_8$ ($\rho_{\mathrm{l}} = 1379\,\mathrm{kg \, m^{-3}}$), projected with SRIM \cite{ziegler_srim_2010} for each $E_{\alpha}$.}
\label{tab:alpha_sphericity_params}
\begin{tabular*}{\textwidth}{l @{\extracolsep{\fill}} lll}
$E_{\alpha}\,\mathrm{[MeV]}$ & $l_{\mathrm{cyl}}\,[\SI{}{\micro\meter}]$ & $R_{\mathrm{s}}\,[\SI{}{\micro\meter}]$ & $t_{\mathrm{s}}\,[\SI{}{\micro\second}]$\\[3pt]\hline\\[-1.5ex]
{5.6}  & 42.6 &  $16.2 \pm 1.2$ & $3.5 \pm 0.3$ \\[5pt] 
{6.1} & 48.3 &  $18.4 \pm 1.4$ & $4.0 \pm 0.3$ \\[5pt]
{7.8} & 70.0 & $26.6 \pm 2.1$ &  $5.7 \pm 0.4$
\end{tabular*}
\end{table}

\subsection{Extraction of acoustic signal}\label{sec:power_extraction}

\begin{figure}
    \subfloat[$\mathcal{P}(t)$ for bubbles nucleated from fluorine recoil-like energy depositions of $\mathcal{T} = 5.5, 50.5, $ and \SI{500.5}{keV}.]{\label{fig:power_nuclear_recs}\includegraphics[scale = 0.3]{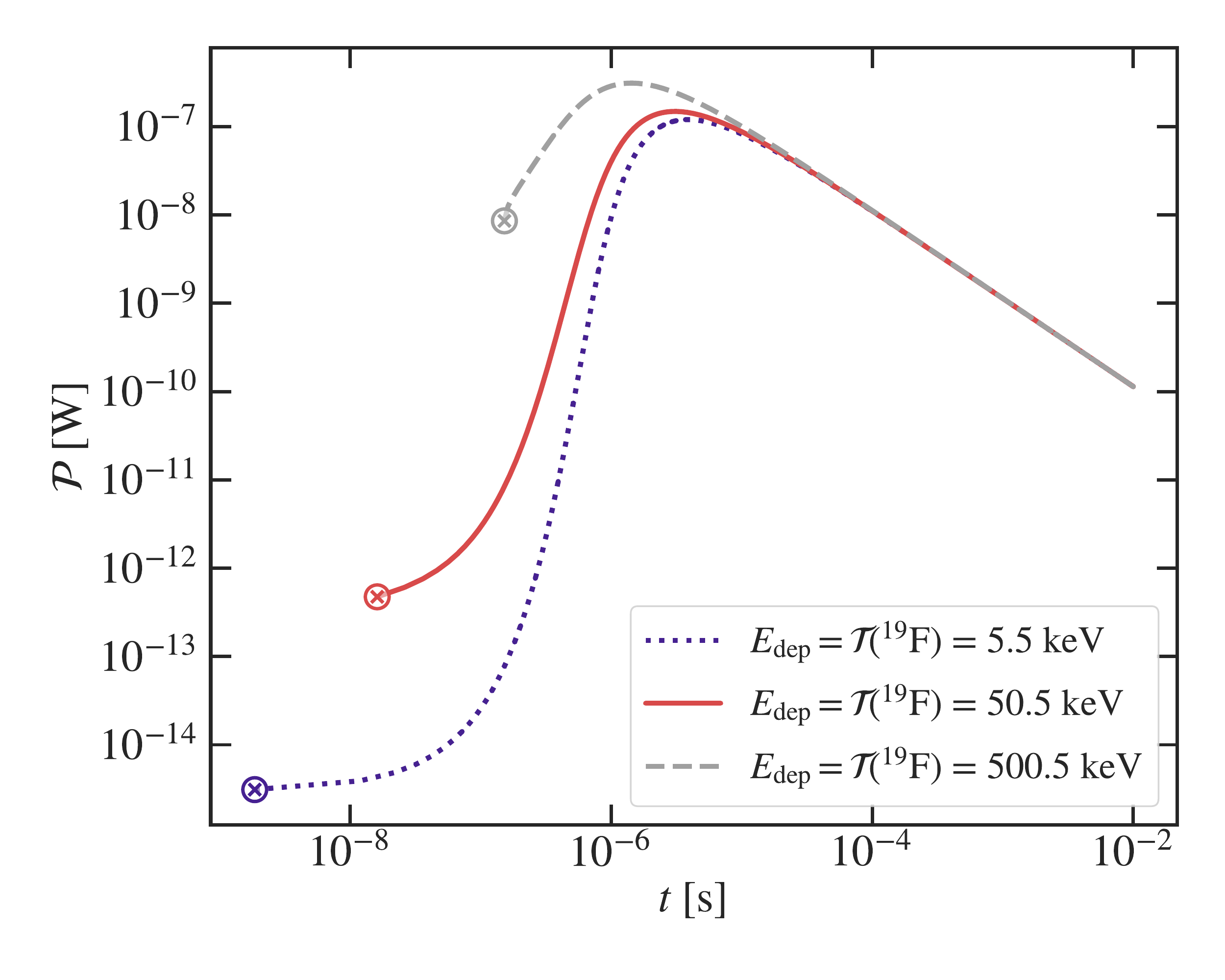}} \\
    \subfloat[$\mathcal{P}(t)$ for bubbles nucleated from $\alpha$-like energy depositions of $E_{\alpha} = 5.6, 6.1, $ and \SI{7.8}{MeV} over the respective track lengths listed in \autoref{tab:alpha_sphericity_params}. ]{\label{fig:power_alphas}\includegraphics[scale = 0.3]{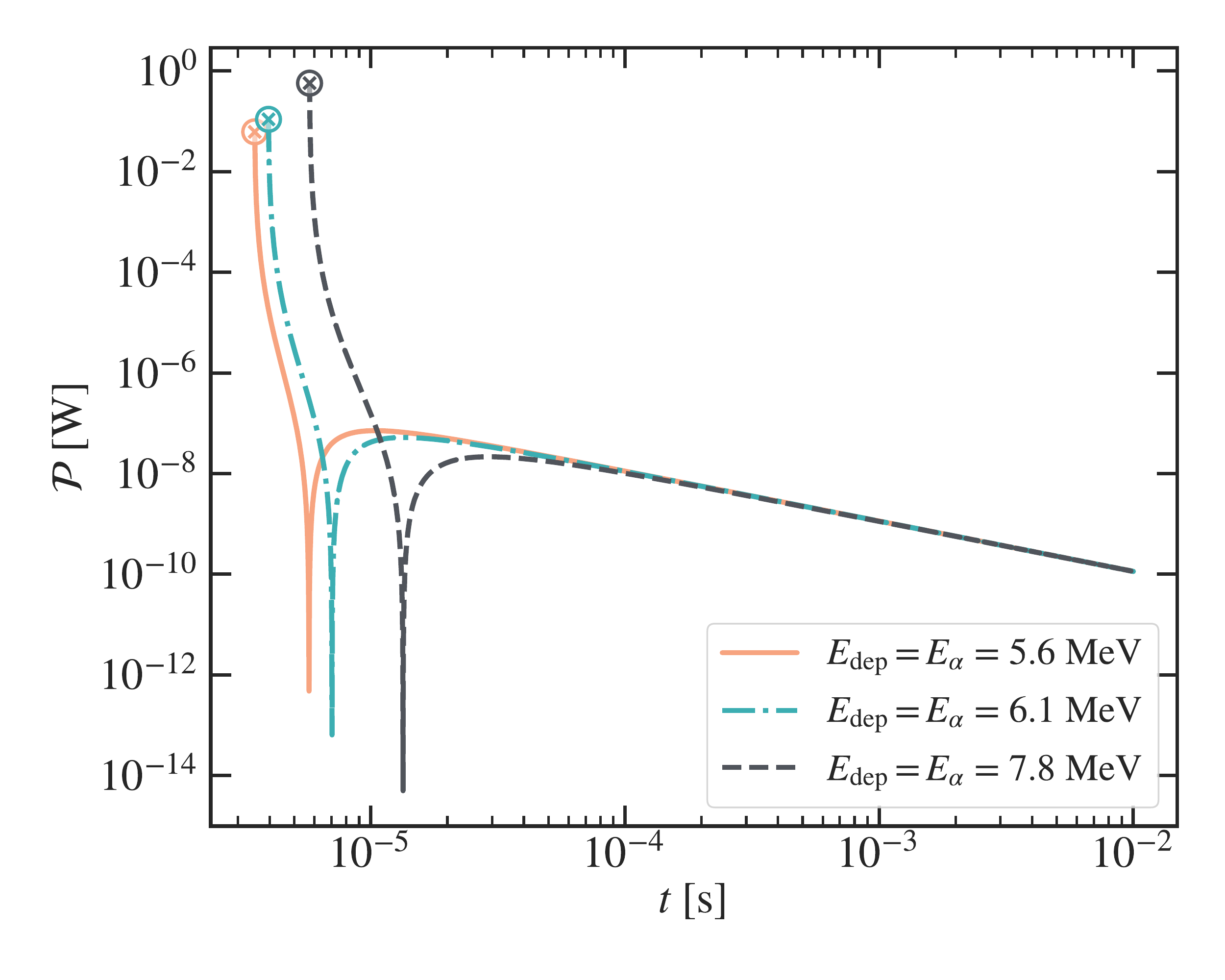}}

    \caption{Acoustic power $\mathcal{P}$ radiated during bubble expansion as a function of time $t$ elapsed after the heat spike. Each $\mathcal{P}(t)$ curve starts at time $t_{\mathrm{s}}$ when the bubble becomes spherical (see \autoref{eq:definition_of_sphericity_time}), which is marked as a circled cross in each plot. The power is evaluated from the bubble radius evolution at $t \geq t_{\mathrm{s}}$ according to \autoref{eq:landau_radiated_power}.}
    \label{fig:power_vs_time}
    
\end{figure}

Using \autoref{eq:full_thermal_DE}, we modeled the evolution of bubbles nucleated as the result of energy depositions by both nuclear recoils (\autoref{sec:mikic_bubble_evolution}) and $\alpha$ particles (\autoref{sec:alpha_bubbles}), starting from the time when these bubbles become spherical. Given $R(t)$ in each case, we follow \cite{landau_fluid_2013} to evaluate the total power radiated in sound waves in all directions:
\begin{equation}
    \label{eq:landau_radiated_power}
    \mathcal{P}(t) = \frac{\rho_{\mathrm{l}} \ddot{V}^2}{4 \pi c},
\end{equation}
where $\ddot{V}$ is the second time derivative of the bubble volume $V(t) \equiv \frac{4}{3} \pi R^3(t)$ and $c \approx 333\,\mathrm{m \, s^{-1}}$ is the speed of sound in the liquid phase of superheated C$_3$F$_8$ at $T =$~\SI{14}{\celsius}. In \autoref{fig:power_vs_time}, we plot the acoustic power evaluated from \autoref{eq:landau_radiated_power} as a function of time for fluorine recoils with energies $\mathcal{T} = 5.5, 50.5,$ and $500.5\,\mathrm{keV}$, with corresponding $l_{\mathrm{cyl}}$ extracted from SRIM \cite{ziegler_srim_2010}, as well as the three $\alpha$-induced bubbles from \autoref{tab:alpha_sphericity_params}. 

\begin{figure}
  \includegraphics[scale=0.3]{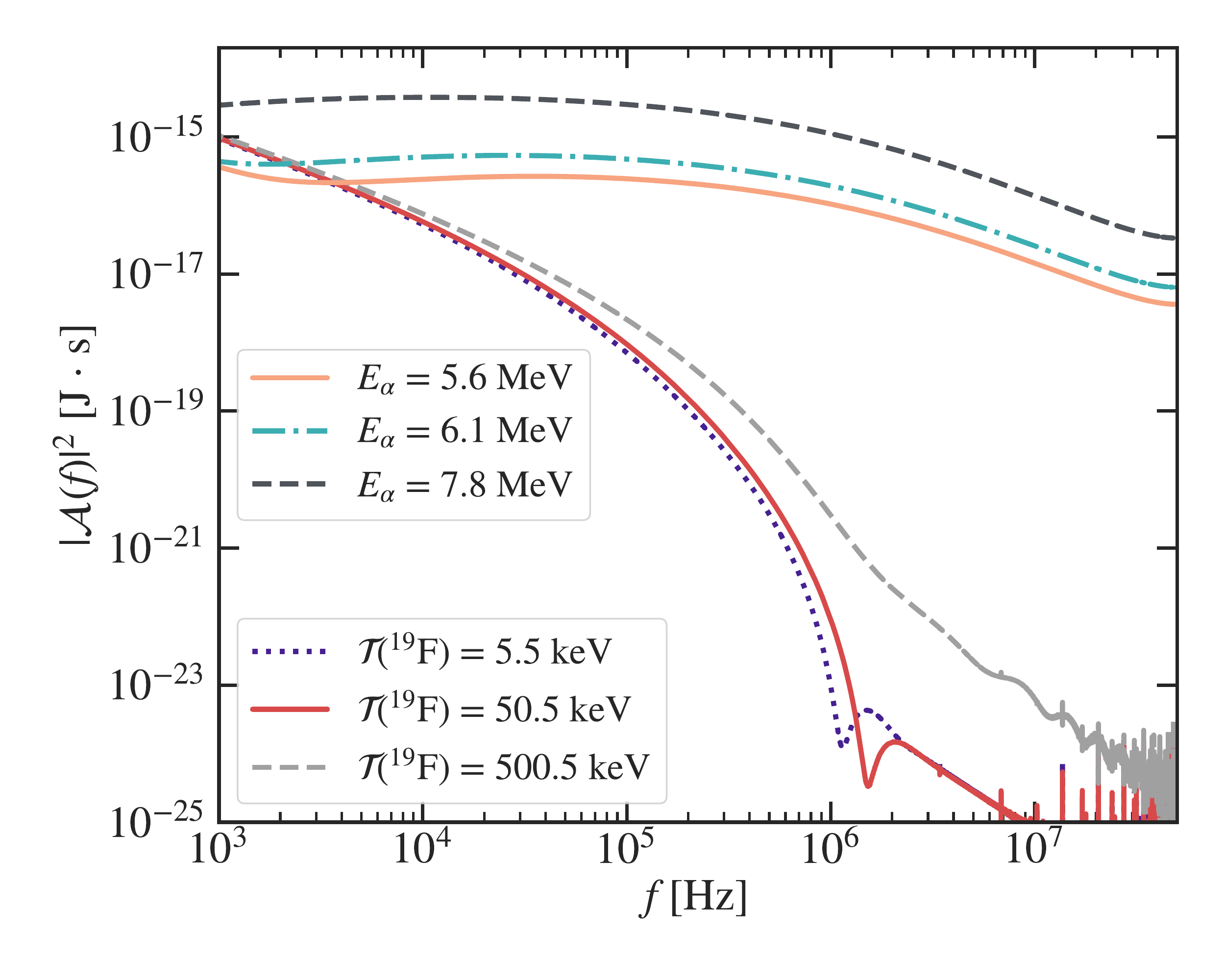}
  \centering
  \caption{Frequency spectra of the acoustic power radiated during expansion of nuclear recoil and $\alpha$-induced bubbles. }
  \label{fig:combined_fourier}
\end{figure}

To find the frequency spectrum of the extracted acoustic signal, we compute discrete Fourier transform (DFT) of the time-dependent acoustic pressure ($p_{\mathrm{ac}} \propto \ddot{V}$). 
\autoref{fig:combined_fourier} gives the resulting frequency spectra for the cases considered in \autoref{fig:power_vs_time}. In this study, we focus on the frequency range of \SIrange[range-phrase=--,range-units=single]{1}{300}{kHz} to compare our distributions of acoustic energies $E_{\mathrm{ac}}$ to those observed in the PICO-60 calibration runs \cite{mitra_pico-60:_2018, amole_dark_2017}. Specifically, we evaluate
\begin{equation}
    E_{\mathrm{ac}}(\SIrange[range-phrase=\text{--},range-units=single]{1}{300}{kHz}) \equiv \int_{ 1\,\mathrm{kHz}}^{300\,\mathrm{kHz}} \left|\mathcal{A}(f) \right|^2 \mathrm{d}f,
\end{equation}
with $A(f)$ being the complex amplitude of the DFT at a frequency $f$ in the returned samples and $\left|\mathcal{A}(f) \right|^2$ giving the respective power spectrum. 

From \autoref{fig:power_nuclear_recs}, we see that at times $t \geq t_{\mathrm{s}}$, only a short interval covers $\mathcal{P}(t)$ monotonically increasing with $t$ for nuclear recoil events. Near $t \sim 1\,\SI{}{\micro\second}$, a maximum in $\mathcal{P}(t)$ is observed, after which the bubble growth slows down in the thermal phase and $\mathcal{P}(t)$ starts to drop. In \autoref{fig:power_alphas}, we find a similar pattern in the acoustic power radiation due to expansion of $\alpha$-induced bubbles. A sudden drop in $\mathcal{P}(t)$ indicates an inflection point characteristic of the times close to $t_{\mathrm{s}}$. At all times prior to $t_{\mathrm{s}}$, the bubbles are not yet spherical and hence the exact $R(t)$ behavior cannot be predicted with the current model. However, using the $R(t)$ data obtained directly from LAMMPS simulations, we found that $\mathcal{P}(t)$ at $t \leq t_{\mathrm{s}}$ is nearly six orders of magnitude smaller than $\mathcal{P}(t)$ at $t \geq t_{\mathrm{s}}$ in the \SIrange[range-phrase=--,range-units=single]{1}{300}{kHz} frequency range, which lets us neglect the contribution from the former.

\section{Generation of acoustic parameter distribution}\label{sec:ap_plot}
In the PICO experiment, a typical data analysis reports the distribution of AP (acoustic parameter) values. The latter are derived directly from the magnitude of the acoustic signal recorded during the expansion of all bubbles observed as single events in the fiducial volume during the run. Neutron-induced nuclear recoils form a low-AP peak in the AP distribution. Expansion of $\alpha$-induced bubbles, which are background events in both calibration and dark matter search runs, is accompanied by radiation of larger acoustic energies. Accordingly, $\alpha$ events feature AP values higher than those characteristic of nuclear recoil (NR) events, and form distinguishable peaks to the right of the latter. The collected data are typically normalized so that center of the NR peak has an AP value of 1. 

In \autoref{sec:power_extraction}, we described a general technique for evaluating the acoustic energy $E_{\mathrm{ac}}$ emitted in \SIrange[range-phrase=--,range-units=single]{1}{300}{kHz} frequency range for both nuclear recoil and $\alpha$-like energy depositions. Additionally, we provided specific examples of $E_{\mathrm{ac}}$ calculation for fluorine recoil energies $\mathcal{T} \in \{5.5, 50.5, 500.5\}\,\mathrm{keV}$ and alpha energies $E_{\alpha} \in \{5.6, 6.1, 7.8\}\,\mathrm{MeV}$. For all cases, a uniform energy deposition along the SRIM-derived mean ranges of F/He ions in C$_3$F$_8$ was assumed. In reality, however, each projected track length value is subject to certain variation $\sigma(l_{\mathrm{cyl}})$ due to collisions, which can be extracted from SRIM as longitudinal straggling \cite{ziegler_srim_2010}. This variation will inevitably translate into a spread in times $t_{\mathrm{s}}$ when the nucleated bubbles become spherical, as well as their radii $R_{\mathrm{s}} = R(t_{\mathrm{s}})$, according to \cref{eq:definition_of_sphericity_time,eq:definition_of_sphericity_radius}. Ultimately, we will obtain a distribution of radiated acoustic energies that will reflect the variations in the length of the energy deposition region $l_{\mathrm{cyl}}$. 

\begin{figure}
  \includegraphics[scale=0.3]{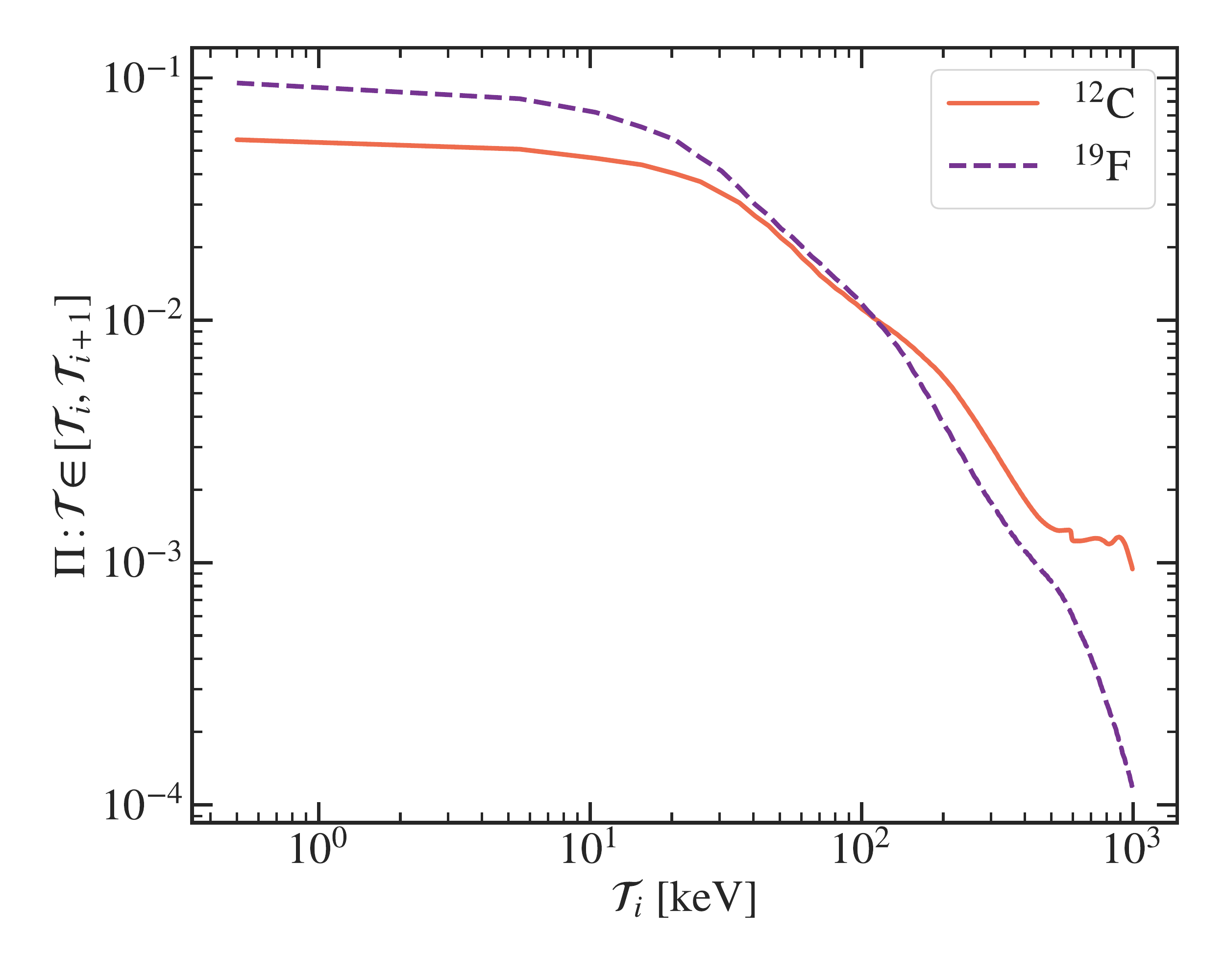}
  \centering
  \caption{Normalized probabilities of C/F recoils having the energy between $\mathcal{T}_i$ and $\mathcal{T}_{i+1} = \mathcal{T}_i + \Delta \mathcal{T}$ ($\Delta \mathcal{T} = 5\,\mathrm{keV}$) as the result of elastic scattering of $^{241}\mathrm{Am}/^{9}\mathrm{Be}$ source neutrons (the spectrum measured in \cite{marsh_high_1995}). For a neutron entering the C$_3$F$_8$ medium, the shown probability curves will be suppressed by factors of 3/11 and 8/11 for C and F nuclei respectively, which we take into account by generating 3,000 carbon recoil-like samples and 8,000 fluorine recoil-like samples when modeling the distributions of radiated acoustic energies. }
  \label{fig:recoil_energy_probabilities}
\end{figure}

For the nuclear recoil energies $\mathcal{T}$, we consider a range from 0.5 to \SI{1000.5}{keV}. Below \SI{0.5}{keV}, the probability of nucleating a stably growing bubble in C$_3$F$_8$ is zero regardless of the energy deposition geometry, which follows the Seitz threshold model \cite{seitz_theory_1958} and was confirmed both experimentally \cite{amole_dark_2017,amole_dark_2019} and through the MD simulations for our LJ fluid. Taking into account recoil energies exceeding $1\,\mathrm{MeV}$ will not contribute much to the resulting distribution of acoustic energies, for the probability of such high $\mathcal{T}$ is very small. We verify this by invoking the kinematics of $^{241}\mathrm{Am}/^{9}\mathrm{Be}$ source neutrons, whose original spectra were measured in e.g.\:\cite{marsh_high_1995} with high precision. These neutrons enter the detector from the outside and propagate through the hydraulic fluid on their way to the target C$_3$F$_8$. By simulating this propagation in Geant4 \cite{agostinelli_geant4simulation_2003} with the geometry of the PICO-40L detector, we obtain a shifted spectrum of $^{241}\mathrm{Am}/^{9}\mathrm{Be}$ neutrons, now directly incident on the target nuclei. We note that an $^{241}\mathrm{Am}/^{9}\mathrm{Be}$ source spectrum closely resembles that of a $^{252}\mathrm{Cf}$ source, which is why we omit a separate discussion of $^{252}\mathrm{Cf}$ neutron kinematics. Convolution of this spectrum with the cross sections of elastic neutron-nucleus scattering \cite{brown_endf/b-viii.0:_2018} gives the distribution of the normalized probabilities of carbon and fluorine recoils shown in \autoref{fig:recoil_energy_probabilities}. We see that recoil energies above \SI{1}{MeV} are less probable than the energies up to \SI{100}{keV} by nearly 2 orders of magnitude, justifying our choice of the recoil energy range.

To generate an AP distribution resembling that obtained experimentally, we consider a sample of 11,000 neutron-induced recoils (among which 3,000 are those of $^{12}$C and 8,000 are those of $^{19}$F) and 5,000 energy depositions by each of the $\alpha$ particle types \eqref{eq:decay_chains}. The NR sample has a well-defined continuous distribution of energies according to \autoref{fig:recoil_energy_probabilities}. Given $N_i$ samples in the recoil energy range from $\mathcal{T}_i$ to $\mathcal{T}_i + \Delta \mathcal{T}$ ($\Delta \mathcal{T} = \SI{5}{keV}$), we can generate a Gaussian distribution of C/F ranges centered at the SRIM-derived mean values for the energy ${T}_i$ and having $\sigma(l_{\mathrm{cyl}})$ as a standard deviation. \cref{eq:definition_of_sphericity_time,eq:definition_of_sphericity_radius} are then used to predict the time $t_{\mathrm{s}}$ when the nucleated bubbles will become spherical and their radii $R_{\mathrm{s}}$ at that point. From here, each NR instance in the distribution will have an associated bubble growth curve $R(t)$ generated from \autoref{eq:full_thermal_DE}. In turn, $R(t)$ gives the acoustic power $\mathcal{P}(t)$ radiated at $t \geq t_{\mathrm{s}}$ (see \autoref{eq:landau_radiated_power}). The acoustic pressure signal is then Fourier-transformed, and its normalized power spectrum is integrated over the frequency range of \SIrange[range-phrase=--,range-units=single]{1}{300}{kHz} to obtain the acoustic energy $E_{\mathrm{ac}}$ emitted at $t \geq t_{\mathrm{s}}$ up to \SI{0.01}{s}. In this way, for each recoil energy $\mathcal{T}_i$ of carbon and fluorine ions, we are able to evaluate the acoustic energy emission at all stages of bubble growth following $t_{\mathrm{s}}$. This part of the calculation results in the low-$E_{\mathrm{ac}}$ nuclear recoil peak in the modeled distribution. The procedure of generating the distribution of acoustic energies radiated during expansion of the $\alpha$-induced bubbles is effectively identical, except only three populations of deposited energies are possible, which we assume to be equally probable.

\autoref{sec:results} presents the distribution of acoustic energies radiated by neutron-induced and $\alpha$-induced bubbles in a PICO-like bubble chamber. The uncertainties on $E_{\mathrm{ac}}$ values that are derived from the uncertainties on the model parameters ($A, B, \gamma, \nu_{\mathrm{l}}, R_{\mathrm{s}}$, and $t_{\mathrm{s}}$) are not taken into account in generating such a distribution, as it would only result in a global shift of all AP values. 

\section{Results and discussion}\label{sec:results}

\begin{figure}[h!]
    \subfloat[Distribution of the modeled AP values for bubbles nucleated by neutron-induced C/F recoils and the three $^{222}\mathrm{Rn}$ decay chain $\alpha$-particle populations \eqref{eq:decay_chains}. The data is normalized so that the Gaussian center of the NR peak has an AP$_{\mathrm{mod}}$ value of 1. The procedure followed to arrive at the AP$_{\mathrm{mod}}$ distribution is described in \autoref{sec:ap_plot}.]{\label{fig:modeled_AP_plot}\includegraphics[scale = 0.3]{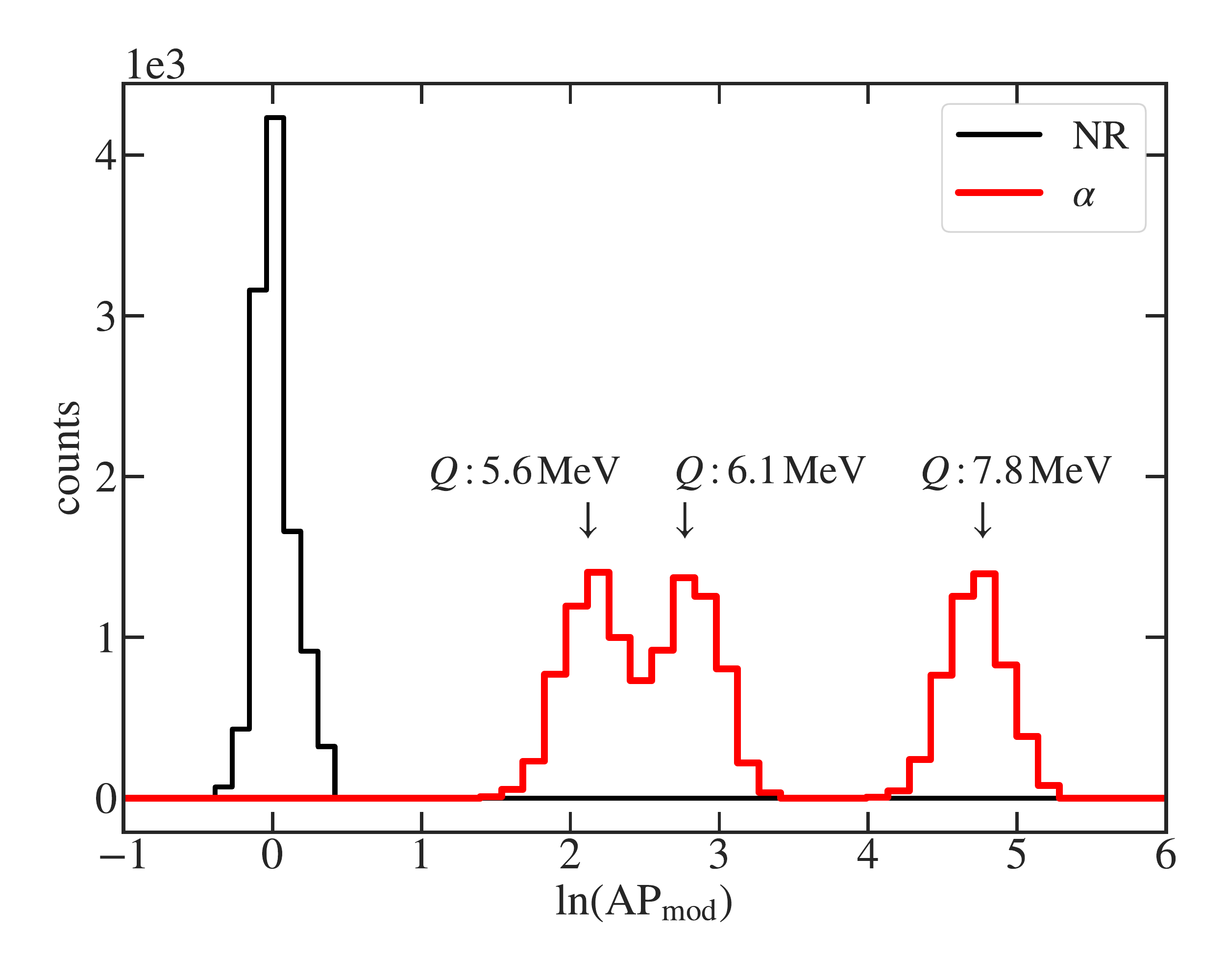}} \\
    \subfloat[AP distribution as obtained from the PICO-60 run at temperature $T_0 = 13.9^{\circ}\mathrm{C}$ and pressure $P_{\mathrm{l}} = $ \SI{30.2}{psia} (\SI{208}{kPa}), corresponding to the bubble nucleation energy threshold of $3.3\,\mathrm{keV}$ \cite{amole_dark_2017, amole_dark_2019}. Both $^{241}\mathrm{Am}/^{9}\mathrm{Be}$ and $^{252}\mathrm{Cf}$ sources were used for neutron calibration. ]{\label{fig:pico60_experimental_AP_plot}\includegraphics[scale = 0.3]{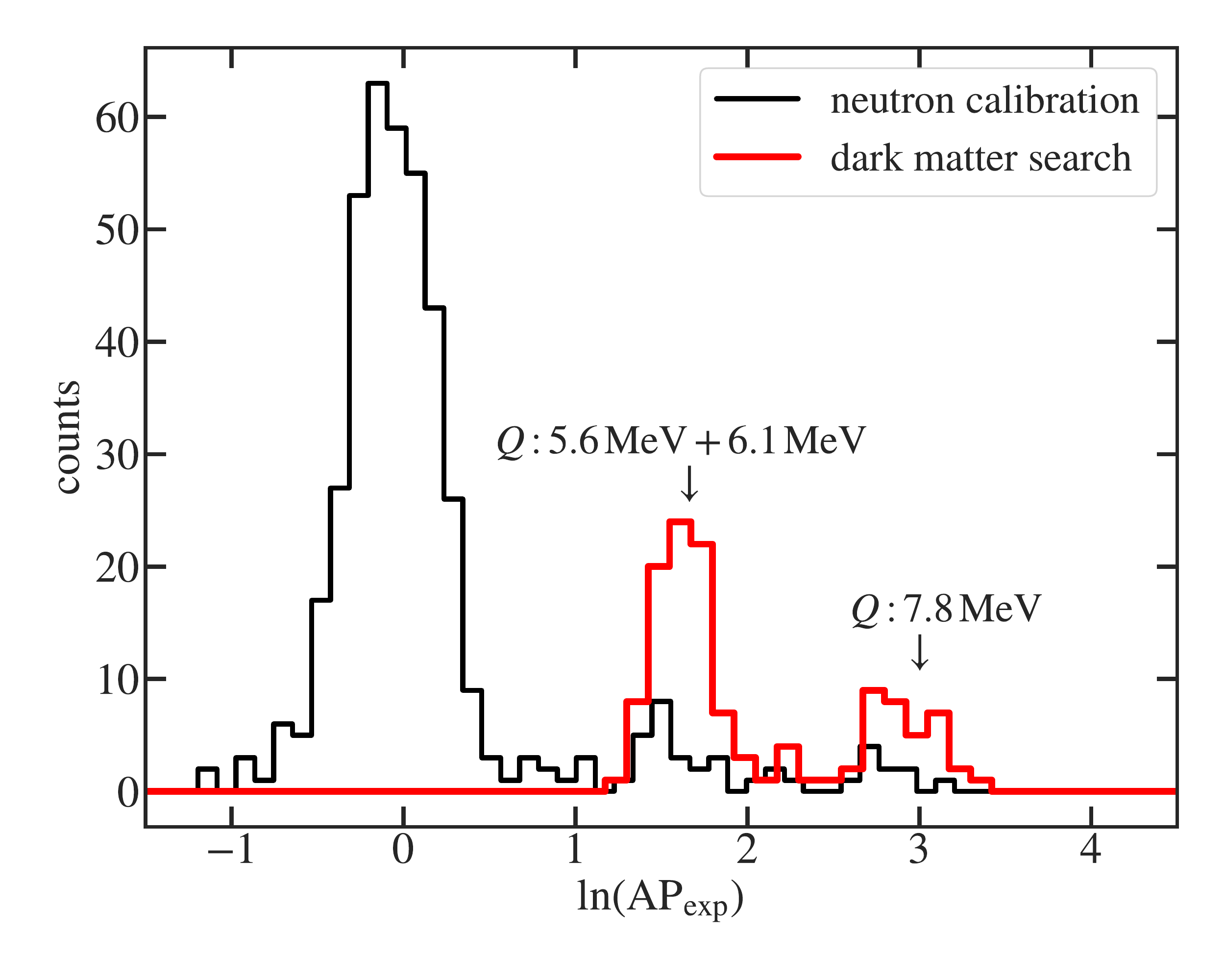}}

    \caption{Comparison of the modeled and the experimental acoustic parameter (AP) distributions. ``NR'' stands for the population of bubbles induced by nuclear recoils, which occur due to elastic scattering of calibration neutrons. The rest of the presented data, which appears under the label ``$\alpha$,'' is the population of bubbles nucleated by $\alpha$ particles from the $^{222}\mathrm{Rn} \to\,^{210}\mathrm{Pb}$ decay chain \eqref{eq:decay_chains}. $Q$-values of the respective decays are listed above each $\alpha$ peak. In deriving AP$_{\mathrm{mod}}$, we assumed energy depositions by the alphas to be equal to these $Q$-values.}
    \label{fig:AP_comparison_fig}
    
\end{figure}

Following renormalization of the simulated acoustic energies by the Gaussian mean of those radiated in nuclear recoil-induced bubble expansions, we obtain the distribution of AP values shown in \autoref{fig:modeled_AP_plot}. We immediately observe that the derived AP pattern provides clear discrimination between neutron-induced nuclear recoils and $\alpha$ particles in  \SIrange[range-phrase=--,range-units=single]{1}{300}{kHz} frequency range. It can therefore be directly compared to the experimental AP distribution from the PICO-60 run \cite{amole_dark_2017, amole_dark_2019} with \SI{52}{kg} of C$_3$F$_8$, which we reproduce in \autoref{fig:pico60_experimental_AP_plot}.

Comparing the modeled and the experimentally obtained AP patterns, we report good qualitative agreement between the two, with larger AP values being a characteristic signature of $\alpha$ particle-induced bubbles in both cases. This outcome confirms that the spherical bubble growth model \eqref{eq:full_thermal_DE} is a plausible description of bubble dynamics for both NR- and $\alpha$-induced bubbles, as it correctly predicts the relative magnitudes of acoustic signals associated with the two bubble nucleation scenarios. The AP-based $\alpha$ particle discrimination used by the PICO collaboration thereby becomes fully motivated by a physical model for the first time. 

Quantitatively, we find differences between the modeled and the simulated AP in both the means of the NR/alpha peaks and the spread of AP values about these means. The discrepancies between the exact positions of the corresponding peaks on \cref{fig:modeled_AP_plot,fig:pico60_experimental_AP_plot} are not surprising. Indeed, in this study, we evaluated acoustic energy emitted in all directions during bubble expansion and converted it to AP$_{\mathrm{mod}}$ directly. In the experimental data, AP$_{\mathrm{exp}}$ is a measure of the acoustic signal magnitude as recorded by the piezoelectric sensors, the response of which is dependent not only on the total energy radiated in sound waves, but also on the exact sound wave propagation characteristics and the accompanying chamber wall vibrations. Several simplifications have been made prior to arriving at the acoustic emission model as well, such as describing C$_3$F$_8$ as a LJ fluid at the stage of molecular dynamics simulations.  We therefore do not expect the AP$_{\mathrm{mod}}$ values to match the AP$_{\mathrm{exp}}$ ones exactly. The question of intrinsically higher NR peak resolution and clearer separation of the \SI{5.6}{MeV} and \SI{6.1}{MeV} alpha peaks is of a similar nature. In our model, only the energies deposited and the statistical variation in the respective ion ranges give rise to a continuum of radiated acoustic energies. In a more realistic scenario, the presence of noise in the recorded signal, limited transducer sensitivity, and the precision of bubble position reconstruction all contribute to the width of each observed peak, along with the aforementioned physical reasons. A full treatment of sound wave propagation in the PICO bubble chambers and instrumental responses to the original acoustic signal would be required to predict the exact smearing effects on the peaks from \autoref{fig:modeled_AP_plot}. Such a study is beyond the scope of the present work and is in progress within the PICO collaboration.

\section{Conclusions}\label{sec:discussion}

In this paper, we have attempted to physically motivate acoustic $\alpha$-particle background discrimination in C$_3$F$_8$-filled PICO bubble chambers. The model that we have constructed builds on predictions for bubble growth in the superheated C$_3$F$_8$-like Lennard-Jones fluid, which we simulated with the methods of molecular dynamics. The bubble evolution was found to depend heavily on initial conditions set by the magnitude and geometry of energy depositions by nuclear recoils and $\alpha$ particles but was otherwise driven by exactly the same physics in the two cases. We have shown that $^{222}\mathrm{Rn} \to\,^{210}\mathrm{Pb}$ chain $\alpha$ particles inevitably nucleate only a single bubble per each $\alpha$ track in the simulated fluid. This has given us the grounds to apply a universal approach in evaluating the acoustic energies radiated during bubble expansion by both $\alpha$-induced and nuclear recoil-induced bubbles, based on the volume change rate of a single bubble. With this simple idea, we were able to reproduce the main empirical finding of \cite{aubin_discrimination_2008}; namely, the stronger acoustic signature of all $\alpha$ events relative to the nuclear recoils.  This feature was found to originate from the larger volume change rate of $\alpha$-induced bubbles at the times when most acoustic energy is emitted. The successful modeling of the acoustic spectrum validates our understanding of the bubble formation and acoustic emission based on the molecular dynamics simulations, applicable to both types of ionizing radiation considered. We thereby come to the first model-based justification for the approach used by PICO to discriminate $\alpha$ background from calibration neutron-induced nuclear recoils, which are kinematically similar to WIMP-induced recoils.

The present study can be further expanded by considering other types of background radiation relevant to underground bubble chambers. In particular, bubble nucleation by electron recoils has been of continued interest. The PICO collaboration has recently discovered that the Seitz model \cite{seitz_theory_1958} for bubble nucleation at moderate superheats only applies to nuclear recoils and alpha interactions \cite{electron_recoil_paper}. Electron recoils, due to the extraordinary suppression at the thermodynamic operation point of dark matter bubble chambers, follow a different model of threshold behavior than that predicted by the Seitz model. Therefore a separate set of molecular dynamics simulations would be required to investigate how the electron recoil bubble nucleation mechanism influences the timeline and magnitude of acoustic emission.\\[10pt]

\section{Acknowledgements}
We wish to acknowledge an earlier study on molecular dynamics simulations of bubble nucleation in direct dark matter search detectors by Denzel et al. \cite{denzel_molecular_2016}, which formed the foundation of the present work. In particular, we thank Philipp Denzel for useful clarifications on their publication and making the developed code publicly available via \href{phdenzel.github.io}{phdenzel.github.io}. The computational resources and support were kindly provided by Compute Canada (\href{www.computecanada.ca}{www.computecanada.ca}), SciNet, and WestGrid. We thank Gabriel Hanna, Bruce Sutherland, and John Beamish of the University of Alberta for helpful discussions on components of this study. We also wish to acknowledge the support of the Natural Sciences and Engineering Research Council of Canada (NSERC) for funding.
\newpage 

\appendix
\counterwithin{figure}{section}

\section{Lennard-Jones potential parameters for C$_3$F$_8$}\label{appendix:appendix_lj_parameters}

In molecular dynamics simulations, the Lennard-Jones (LJ) potential \cite{jones_determination_1924} is a common way to describe simple interatomic interactions pair-by-pair. The LJ potential is defined through the characteristic energy (potential well depth) $\epsilon$ and the characteristic length (effective atomic diameter) $\sigma$, which are the units used within MD simulations. Such a potential is typically cut off at a certain critical distance to reduce computational costs of a simulation. To minimize the error of such a truncation, the interaction force is shifted so that it goes to zero continuously, resulting in a truncated shifted-force Lennard-Jones (TSF-LJ) potential \cite{toxvaerd_communication:_2011}. Following \cite{denzel_molecular_2016,diemand_direct_2014}, we choose the cutoff radius to be $2.5 \sigma$, which ensures the deviation from the full-tail $u_{\mathrm{LJ}}$(r) to be smaller than $1.6\%$ \cite{toxvaerd_communication:_2011}.

Since we aim to compare the results of the present study with the acoustic data from the PICO-60 run at $\sim$\SI{14}{\celsius} \cite{amole_dark_2017}, it is important to reproduce the major thermodynamic properties of C$_3$F$_8$ at this particular temperature, corresponding to $T_0 = 0.778\,\epsilon/k_{\mathrm{B}}$ in LJ units. For that reason, we repeated the standard liquid-vapor interface experiment (see e.g. \cite{diemand_direct_2014}) for $0.778\,\epsilon/k_{\mathrm{B}}$ to extract equilibrium pressure $P_{\mathrm{eq}}$, liquid/vapor density $\rho_{\mathrm{l}}/\rho_{\mathrm{v}}$, and surface tension $\gamma$ in LJ units ($\epsilon \, \sigma^{-3} $, $m \, \sigma^{-3}$, and $\epsilon \, \sigma^{-2}$ respectively, where $m$ = \SI{188.02}{u} for C$_3$F$_8$) at saturation. We then used the corresponding real values of these quantities from the REFPROP database \cite{LEMMON-RP91} to deduce the suitable value of $\sigma$, with $\epsilon$ fixed at \SI{0.0318}{eV} as discussed in \autoref{sec:simsetup}. We started the experiment by creating a simulation box with length and width $L_x = L_y = 200\sigma$ and height $L_z = 100\sigma$, to which we applied periodic boundary conditions, and placing a liquid slab in the middle of the box. The slab initially contained 1,240,000 ``atoms'' on a lattice, taking the whole $XY$-plane of the simulation box and $30 \sigma$ in $z$ direction. The atoms were assigned Maxwell-distributed velocities corresponding to $T_0$. To preliminarily randomize their positions, we performed a microcanonical (\textit{NVE}: constant particle number, volume, and energy) ensemble simulation for 100,000 time steps. A single time step was set equal to $0.0025 \tau$, where $\tau = \sigma \sqrt{m/\epsilon} \approx \SI{4.172}{ps}$. In each run, we used a Langevin thermostat to control the temperature. After we obtained such stable liquid conditions, we allowed the system to evolve for 2,000,000 time steps as an \textit{NVT} (constant particle number, volume, and temperature) ensemble. The purpose of the first 1,000,000 time steps was to obtain two well-equilibrated liquid-vapor interfaces, below and above the liquid slab, after which we were able to take the measurements $P_{\mathrm{eq}}$ and $\rho_{\mathrm{l}}/\rho_{\mathrm{v}}$ in the subsequent 1,000,000 time steps. The time averaging of these quantities was performed 20 times in 50,000-time-step-long intervals, which ultimately gave the $1\sigma$ scatter for the global weighted average \cite{diemand_direct_2014}. The equilibrium pressure was evaluated as the spatial average of the normal ($z$) component of the pressure tensor \cite{walton_pressure_1983}. To obtain values for liquid and vapor densities, we produced a density profile of the system over the $z$-axis, with $\rho(z)$ spatially averaged over chunks of $0.5 \sigma$ height. The hyperbolic tangent function \cite{chapela_computer_1977} was then used to fit the time-averaged $\rho(z)$ dependence and extract the liquid and vapor densities as fit parameters. Finally, to compute surface tensions, we followed the Kirkwood-Buff approach \cite{kirkwood_statistical_1949}. \cref{tab:liq_slab_measurements} presents the results of the $P_{\mathrm{eq}}$, $\rho_{\mathrm{v}}$, $\rho_{\mathrm{l}}$, and $\gamma$ measurements in LJ units; also listed are the respective real values of these quantities extracted from REFPROP database \cite{LEMMON-RP91} in SI units. We observe that there is not a single value of $\sigma$ that would reconcile all of the measurements simultaneously. As discussed in \autoref{sec:simsetup}, we pick $\sigma \approx 0.533\,\si{nm}$ to approximate ``intermolecular'' interactions within C$_3$F$_8$ with the Lennard-Jones potential. This matches the simulation extracted liquid density at \SI{14}{\celsius} with the corresponding REFPROP value and introduces $\lesssim 30\%$ errors on other quantities listed in \autoref{tab:liq_slab_measurements}.

\begin{table}
\caption{Liquid slab experiment results: MD measurements of surface tension and liquid/vapor densities at saturation ($T = \SI{14}{\celsius}$) in LJ units as compared to REFPROP values in SI units. For each quantity, the last column shows $\sigma$ that makes the measured LJ value match the REFPROP-predicted value. The uncertainties on REFPROP values are derived from the uncertainties on simulation-extracted temperatures for each phase ($T_{\mathrm{l}} = 13.95 \pm 0.05\,^{\circ}\mathrm{C}$, $T_{\mathrm{v}} = 14.4 \pm 0.5\,^{\circ}\mathrm{C}$).}
\label{tab:liq_slab_measurements}
\begin{tabular}{l|lll}
~ & MD simulation & REFPROP & $\sigma [\si{nm}]$\\[3pt]\hline\\[-1.5ex]
$\gamma$  & $0.201 \pm 0.001\,{\frac{\epsilon}{\sigma^2}}$ &  $4.82 \pm 0.05\,\mathrm{\frac{mN}{m}}$ & $0.461 \pm 0.003$ \\[5pt] 
$\rho_{\mathrm{l}}$ & $0.6715 \pm 0.0001\,{\frac{m}{\sigma^3}}$ &  $1385.1 \pm 0.3\, \mathrm{\frac{kg}{m^3}}$ & $0.53293 \pm 0.00005$ \\[5pt]
$\rho_{\mathrm{v}}$ & $0.03983 \pm 0.00009\,{\frac{m}{\sigma^3}}$ & $62 \pm 1\, \mathrm{\frac{kg}{m^3}}$ &  $0.586 \pm 0.003$\\[5pt]
$P_{\mathrm{eq}}$ & $0.02240 \pm 0.00003 {\frac{\epsilon}{\sigma^3}}$ & $6.5 \pm 0.1\,\mathrm{bar}$ &  $0.560 \pm 0.003$
\end{tabular}
\end{table}

\section{Heat spike simulation procedure}\label{appendix:heat_spike_simulation}

For the bubble nucleation simulation, we chose a cubic box extending from $-\frac{L_i}{2}$ to $\frac{L_i}{2}$ in each dimension ($i = \overline{x,y,z}$), with $L_i \approx 363.2\sigma$. This box was filled with 32,000,000 atoms on a face-centered cubic lattice, corresponding to $\rho_{\mathrm{l}} = 0.668 \frac{m}{\sigma^3}$. The latter value was chosen so as to reproduce the density of superheated C$_3$F$_8$ at $T_0 = $ \SI{14}{\celsius}, $P_{\mathrm{l}}$ = \SI{207}{kPa}, which is equal to 1379 $\mathrm{\frac{kg}{m^3}}$ \cite{LEMMON-RP91}. Periodic boundary conditions were applied in all three directions. Following the approach of \cite{denzel_molecular_2016} to bring an LJ fluid to a superheated metastable state, we started our system at a temperature exceeding $T_0$, namely $0.8\,\epsilon/k_{\mathrm{B}}$ (\SI{22}{\celsius}). This gave us stable liquid conditions, which were retained for 10,000 time steps in an \textit{NVE} ensemble to equilibrate the system. The next step was to linearly lower the temperature to the target $0.778\,\epsilon/k_{\mathrm{B}}$ over the time interval of 15,000 steps, which was again done in an \textit{NVE} ensemble. The fixed liquid density was now too low for $0.778\,\epsilon/k_{\mathrm{B}}$. Thus, the liquid became superheated. To equilibrate the system at these new conditions, it was evolved for another 30,000 time steps under \textit{NVE} constraints, with temperature fixed for the first 15,000 steps.

Following the instant of a sufficiently energetic heat spike in the prepared superheated fluid, the approach of Denzel et al. \cite{denzel_molecular_2016} was to integrate the system in an \textit{NVE} ensemble until the pressure in the tiny liquid volume ($\approx [\SI{162}{nm}]^3$) increased significantly. This pressure rise slowed the bubble growth down even on a ns scale, which would not have been a realistic description for the early stages of bubble expansion in any PICO-scale bubble chamber. One way to avoid this complication would be to simulate a much larger liquid volume, naturally resulting in increased computational costs. For this reason, we chose to deviate from the method described in \cite{denzel_molecular_2016} by switching from an \textit{NVE} to an \textit{NPT} ensemble for the post-spike part of the simulation. In \textit{NPT} integration, the overall particle number, pressure, and temperature are preserved, which reproduces typical conditions of a large-scale bubble chamber at the stages of bubble expansion when most acoustic energy is emitted.

For $\alpha$-induced bubble nucleation simulations, we set up a simulation box with $L_x = 1888.6\sigma \approx 1\,\SI{}{\micro\meter}$ and $L_z = L_y = 159.8\sigma \approx 85\,\mathrm{nm}$ in LAMMPS. As before, we required $\rho_{\mathrm{l}} = 0.668\,m \, \sigma^{-3} = 1379\,\mathrm{kg \, m^{-3}}$ and $T = 0.778\,\epsilon/k_{\mathrm{B}} = 14^{\circ}\mathrm{C}$. To superheat the liquid, we followed the same procedure as in the nuclear recoil case. However, several modifications to the energy deposition region geometry were made. We observed that if the cylinder radius $r_{\mathrm{cyl}}$ was set to $2\sigma$ as in the nuclear recoil simulation, the \textit{NPT} integrators failed to maintain the requested $0.778\,\epsilon/k_{\mathrm{B}}$ temperature for the whole system due to overly high temperature within the cylinder. To overcome this issue, we enlarged the radius to $3\sigma$. The cylinder length was set equal to $L_x$.

\section{Bubble surface tracking}\label{appendix:surface_tracking_appendix}

Throughout the MD simulations in LAMMPS, we made use of a grid of cubic cells of size $[4\sigma]^3$ and computed the numbers of atoms falling within these cells as averaged over $1000$ time steps ($\approx \SI{0.01}{ns}$) intervals. After each such interval, we saved a slice of the resulting spatial density distributions into a text file. With the energy deposition axis parallel to the $x$-axis and centered at $y = 0$, $z = 0$, we took the slice between $-2\sigma \leq z \leq 2\sigma$. This allowed us to record the 2D evolution of the post-spike densities in the $XY$ plane.

After that, we applied the \texttt{measure.find\_contours} function of the \texttt{scikit-image} Python module \cite{lorensen_marching_1987,walt_scikit-image:_2014} to extract the bubble contour at each moment in time. This function acts upon the 2D density data and returns discrete sets of points ${x_i, y_i}$ for all the contours where $\rho_{\mathrm{l}}(x_i,y_i) = c$ holds. With a suitable choice of $c$, one of the contours found with this method is that of the vapor-liquid boundary. Our tests showed that choosing the longest contour among those extracted is a perfectly robust method to pick up the bubble ``surface'' boundary if we require $c = 0.4~m\,\sigma^{-3}$ $\approx $ \SI{830}{kg}$\, \mathrm{m^{-3}}$. This method of bubble surface tracking was developed as an alternative to recursive linking of vapor cells used in \cite{denzel_molecular_2016,diemand_direct_2014}. In this study, we are only interested in the evolution of a single bubble, which is why identification of vapor regions in the rest of the liquid volume as in \cite{denzel_molecular_2016, diemand_direct_2014} would be redundant.

\section{Transition from nonspherical to spherical bubble shape and further dynamics}\label{appendix:sphericity_radii_and_time_relations}

\begin{figure}[H]
  \includegraphics[scale=0.3]{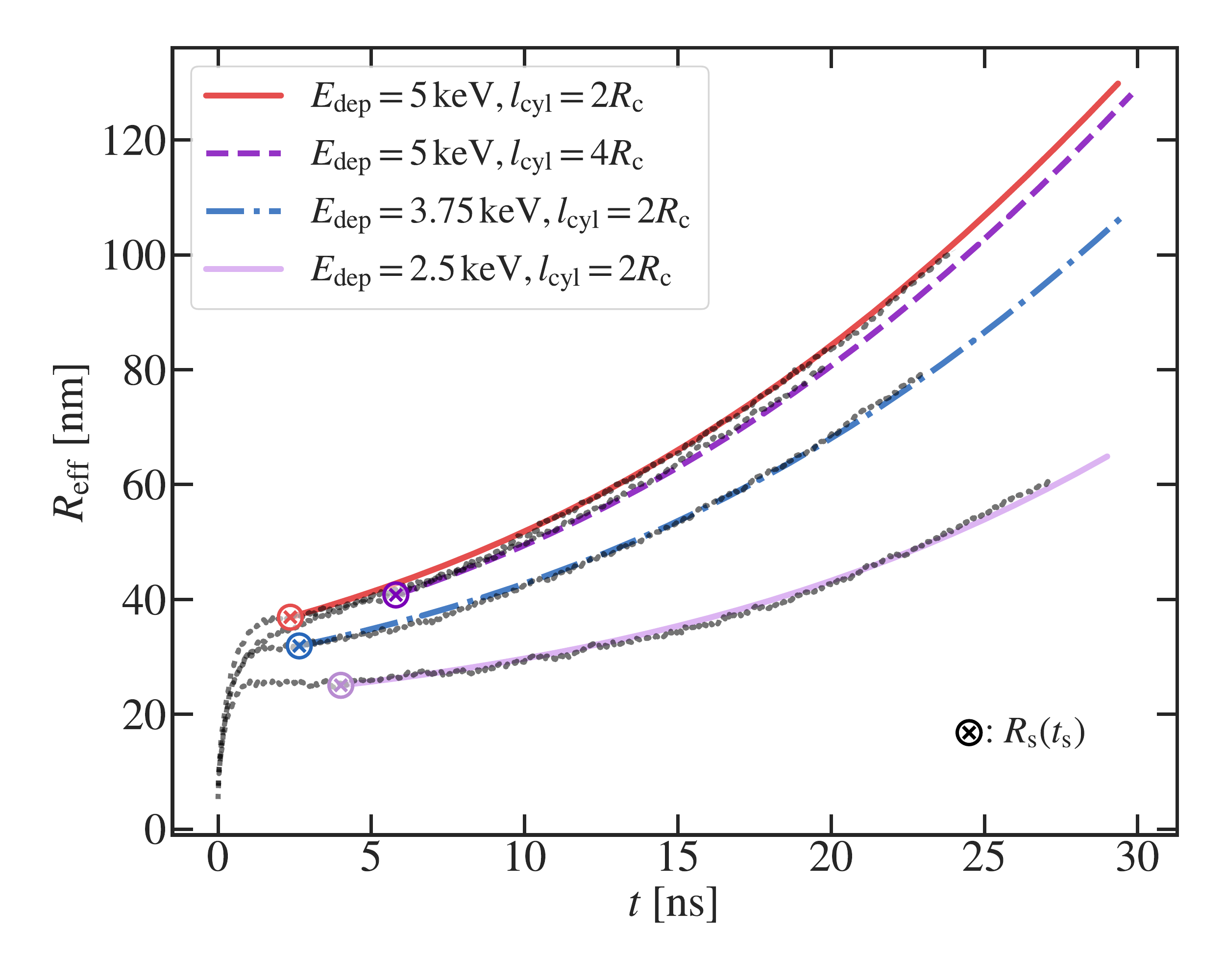}
  \centering
  \caption{The results of simultaneous fitting of the numerical solution to \autoref{eq:full_thermal_DE} to the bubble growth curves obtained from four LAMMPS simulations of bubble nucleation. The cross-shaped markers represent the times when the bubbles become spherical and their respective sizes. These are the starting points of the data portions to be fitted.}
  \label{fig:simultaneous_fitting}
\end{figure}

The differential form of \autoref{eq:full_thermal_DE} is universal for all spherical bubbles irrespective of how they were nucleated. Parameters $A, B, \gamma$, and $\nu_{\mathrm{l}}$ in \autoref{eq:full_thermal_DE} are also shared between the different bubble expansions in the same liquid. We may therefore fit a few of the simulated bubble growth histories $R(t)$ for these parameters and use the best-fitting values later on to predict the dynamics of any bubble expanding within our LJ fluid. For this purpose, we use the bubble evolution data from four simulations mimicking energy depositions by nuclear recoils. The summary of these simulations is given in \autoref{fig:simultaneous_fitting}. We restrict the fit to those parts of the data where the bubbles are already spherical. To avoid overfitting, we find it advantageous to fix $A$ at $17.7\,\mathrm{m\,s^{-1}}$, which is the value we predict from \autoref{eq:definition_of_A} and \autoref{tab:bubble_growth_parameters} for C$_3$F$_8$. We let $B, \gamma$, and $\nu_{\mathrm{l}}$ be free parameters that are shared between the four data sets we described, and arrive at the results in \autoref{tab:bubble_growth_parameters}.

While these parameters are universal for any bubble nucleated within the LJ fluid, the initial conditions for the spherical bubble growth equation \eqref{eq:full_thermal_DE} vary with deposited energy and ion track length. This is reflected in \cref{fig:sphericity_tau,fig:sphericity_radius}. The time $t_\mathrm{s}$ of transition to bubble sphericity has an apparent dependence only on the length of the energy deposition region. The bubble radius $R_\mathrm{s}$ at $t_\mathrm{s}$ is also affected by the energy deposited. We combine these observations in \cref{eq:definition_of_sphericity_time,eq:definition_of_sphericity_radius}.

\noindent

\begin{minipage}[t!]{1\linewidth} 
\begin{figure}[H]
\subfloat[Dependence of $t_{\mathrm{s}}$ on the nucleus track length $l_{\mathrm{cyl}}$ for a fixed $\frac{E_{\mathrm{dep}}}{l_{\mathrm{cyl}}} = \frac{5\,\mathrm{keV}}{4R_{\mathrm{c}}} \approx 57.8\,\mathrm{\frac{keV}{\SI{}{\micro\meter}}}$. We drop the $-0.9 \pm 0.4\,\mathrm{ns}$ intercept as it will become less significant for larger track lengths. 
 ]{\label{fig:spher_tau_on_lcyl}\includegraphics[scale = 0.3]{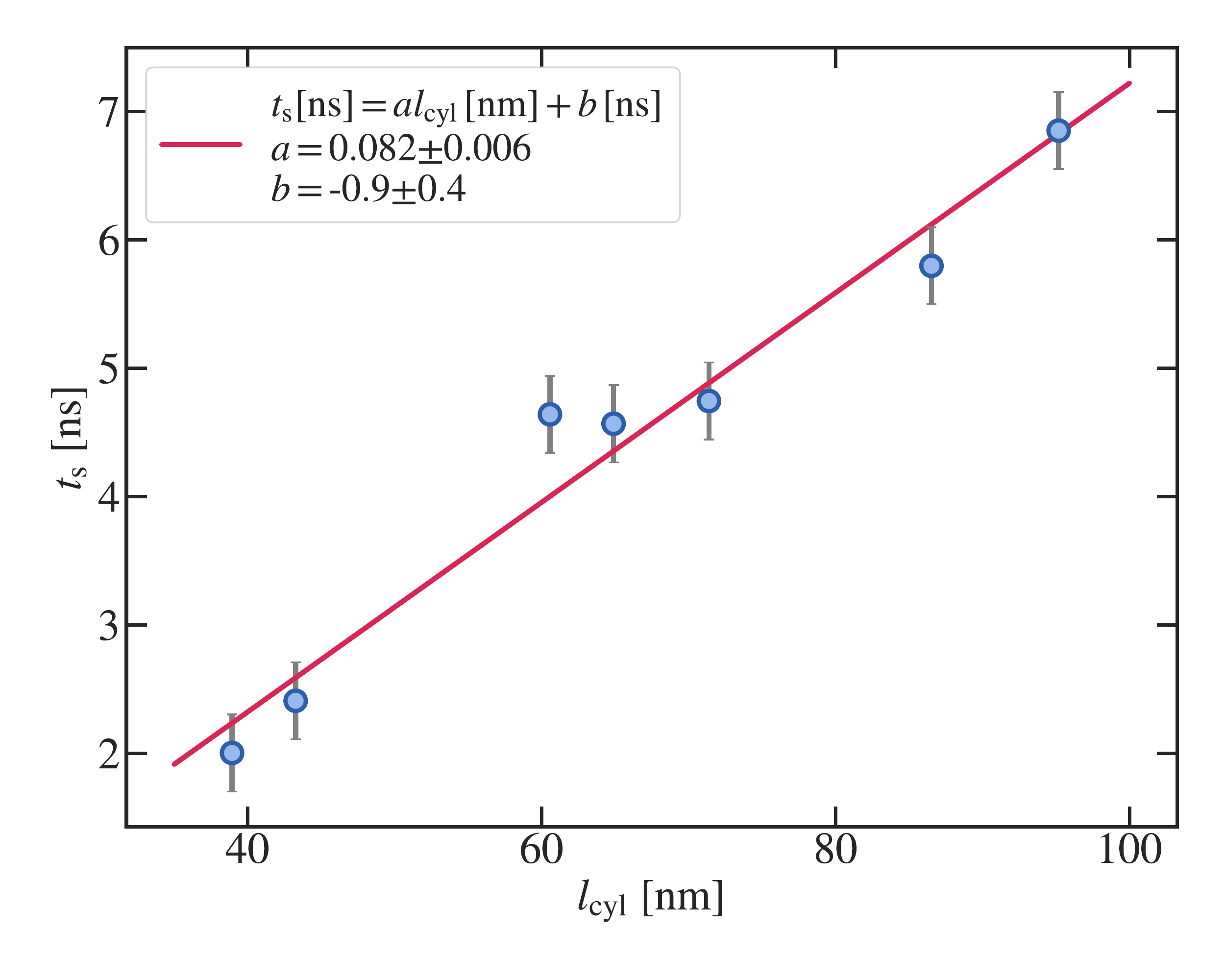}} \\
    \subfloat[Dependence of $t_{\mathrm{s}}$ on the energy $E_{\mathrm{dep}}$ deposited along fixed $l_{\mathrm{cyl}} = 2R_{\mathrm{c}}$. ]{\label{fig:spher_tau_on_Edep}\includegraphics[scale = 0.3]{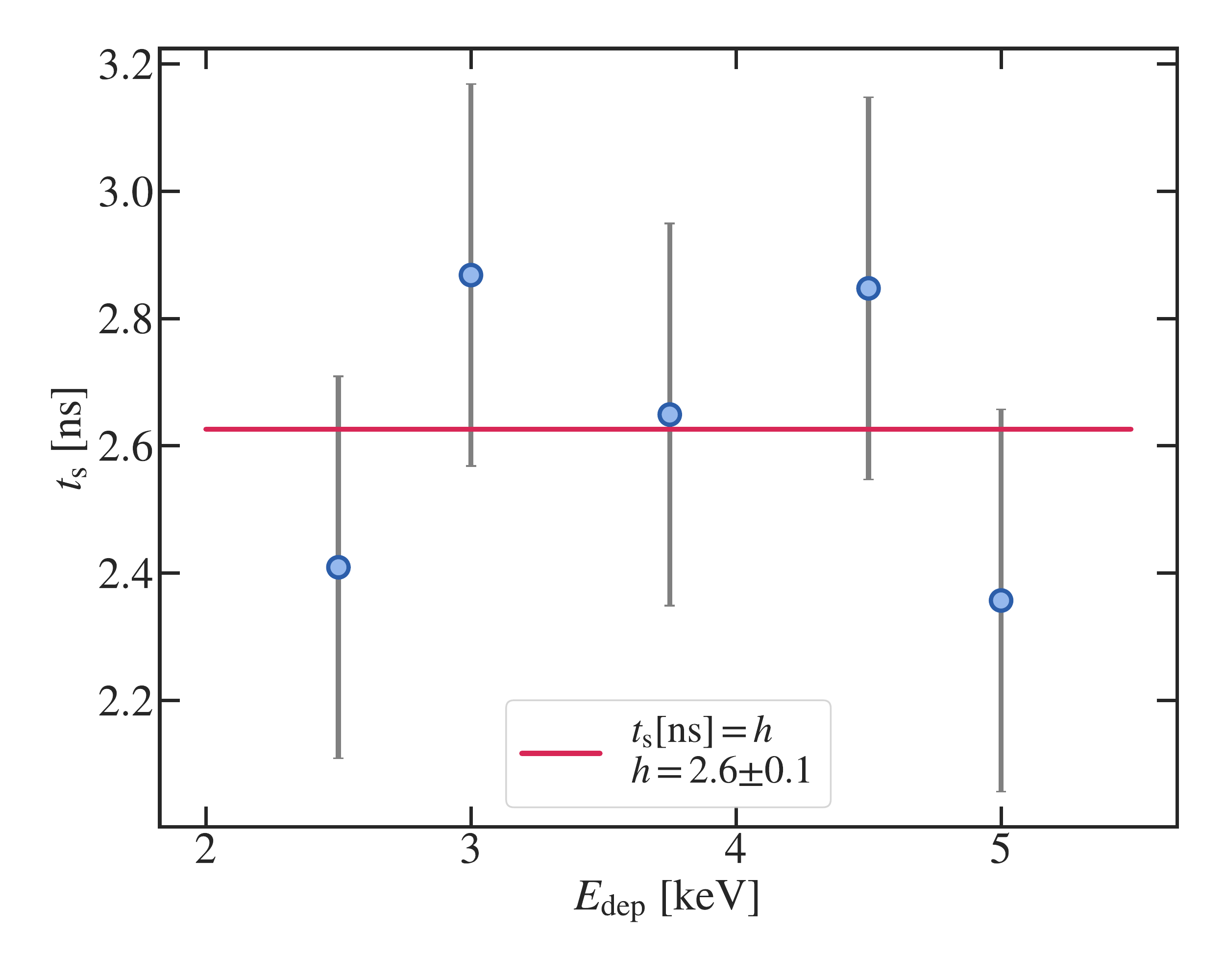}}
    \caption{Variation of the time $t_{\mathrm{s}}$ elapsed after the heat spike when the bubble eccentricity reaches its first minimum (see \autoref{fig:eccentricity_vs_time}) with the energy deposition parameters $E_{\mathrm{dep}}$ and $l_{\mathrm{cyl}}$. The error bars represent the statistical uncertainty on $t_{\mathrm{s}}$ measured with different random seeds.}
    \label{fig:sphericity_tau}
    
\end{figure}
\end{minipage}

\begin{minipage}[t!]{1\linewidth} 
\begin{figure}[H]
    \subfloat[Dependence of $R_{\mathrm{s}}$ on the length of the energy deposition region $l_{\mathrm{cyl}}$ (see \autoref{fig:eccentricity_vs_time}) for a fixed $\frac{E_{\mathrm{dep}}}{l_{\mathrm{cyl}}} = \frac{5\,\mathrm{keV}}{4R_{\mathrm{c}}} \approx 57.8\,\mathrm{\frac{keV}{\SI{}{\micro\meter}}}$. ]{\label{fig:spher_radius_on_lcyl}\includegraphics[scale = 0.3]{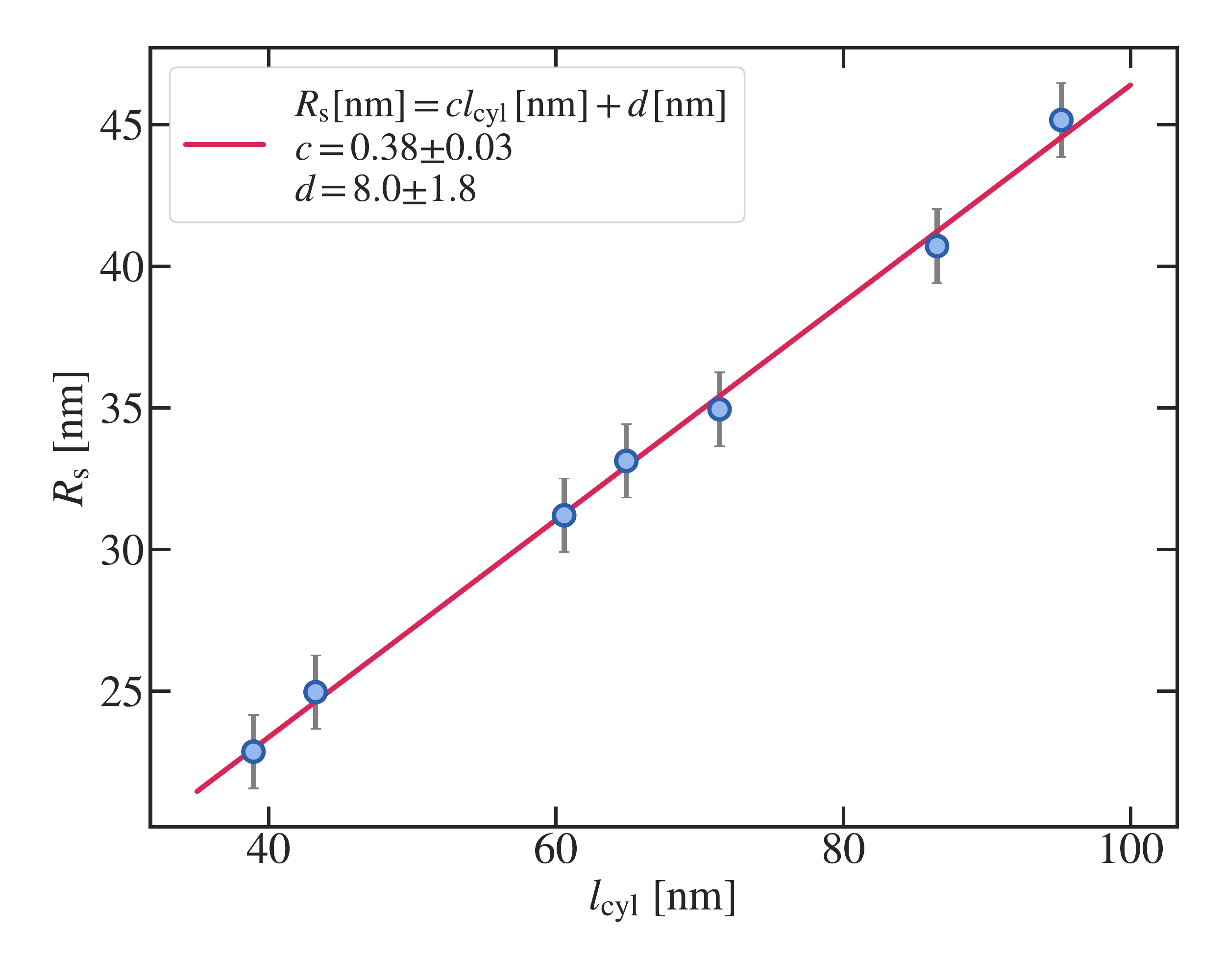}} \\
    \subfloat[Dependence of $R_{\mathrm{s}}$ on the energy $E_{\mathrm{dep}}$ deposited along fixed $l_{\mathrm{cyl}} = 2R_{\mathrm{c}}$. ]{\label{fig:spher_radius_on_Edep}\includegraphics[scale = 0.3]{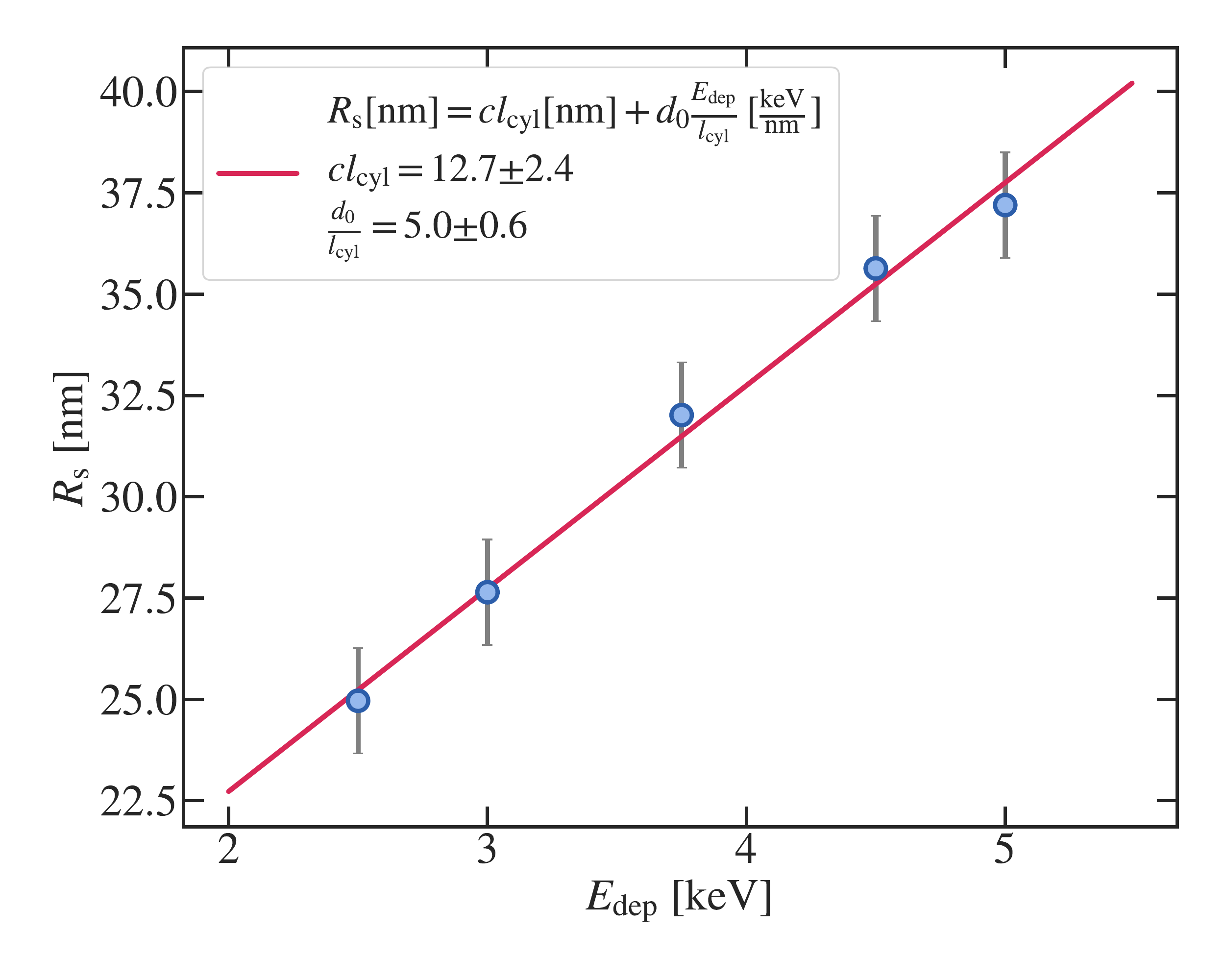}}
    \caption{Variation of the effective bubble radius $R_{\mathrm{s}}$ at the first minimum in the bubble eccentricity (see \autoref{fig:eccentricity_vs_time}) with the energy deposition parameters $E_{\mathrm{dep}}$ and $l_{\mathrm{cyl}}$. The error bars correspond to the statistical uncertainty on $R_{\mathrm{s}}$ measured with different random seeds and the systematic uncertainty equal to a half of the grid cell width ($2\sigma \approx 1\,\mathrm{nm}$). }
    \label{fig:sphericity_radius}
    
\end{figure}
\end{minipage}

\newpage

\bibliographystyle{apsrev.bst}

\begin{thebibliography}{31}
\expandafter\ifx\csname natexlab\endcsname\relax\def\natexlab#1{#1}\fi
\expandafter\ifx\csname bibnamefont\endcsname\relax
  \def\bibnamefont#1{#1}\fi
\expandafter\ifx\csname bibfnamefont\endcsname\relax
  \def\bibfnamefont#1{#1}\fi
\expandafter\ifx\csname citenamefont\endcsname\relax
  \def\citenamefont#1{#1}\fi
\expandafter\ifx\csname url\endcsname\relax
  \def\url#1{\texttt{#1}}\fi
\expandafter\ifx\csname urlprefix\endcsname\relax\def\urlprefix{URL }\fi
\providecommand{\bibinfo}[2]{#2}
\providecommand{\eprint}[2][]{\url{#2}}

\bibitem[{\citenamefont{Glaser and Rahm}(1955)}]{glaser_characteristics_1955}
\bibinfo{author}{\bibfnamefont{D.~A.} \bibnamefont{Glaser}} \bibnamefont{and}
  \bibinfo{author}{\bibfnamefont{D.~C.} \bibnamefont{Rahm}},
  \bibinfo{journal}{Phys. Rev.} \textbf{\bibinfo{volume}{97}},
  \bibinfo{pages}{474} (\bibinfo{year}{1955}).

\bibitem[{\citenamefont{Amole et~al.}(2017)}]{amole_dark_2017}
\bibinfo{author}{\bibfnamefont{C.}~\bibnamefont{Amole}} \bibnamefont{et~al.}
  (\bibinfo{collaboration}{PICO Collaboration}), \bibinfo{journal}{Phys. Rev.
  Lett.} \textbf{\bibinfo{volume}{118}}, \bibinfo{pages}{251301}
  (\bibinfo{year}{2017}).

\bibitem[{\citenamefont{Amole et~al.}(2019{\natexlab{a}})}]{amole_dark_2019}
\bibinfo{author}{\bibfnamefont{C.}~\bibnamefont{Amole}} \bibnamefont{et~al.}
  (\bibinfo{collaboration}{PICO Collaboration}),
  \bibinfo{journal}{arXiv:1902.04031 [astro-ph.CO]}
  (\bibinfo{year}{2019}{\natexlab{a}}).

\bibitem[{\citenamefont{Mitra}(2018)}]{mitra_pico-60:_2018}
\bibinfo{author}{\bibfnamefont{P.}~\bibnamefont{Mitra}}, Ph.D. thesis,
  \bibinfo{school}{University of Alberta} (\bibinfo{year}{2018}).

\bibitem[{\citenamefont{Aubin et~al.}(2008)\citenamefont{Aubin, Auger, Genest,
  Giroux, Gornea, Faust, Leroy, Lessard, Martin, Morlat
  et~al.}}]{aubin_discrimination_2008}
\bibinfo{author}{\bibfnamefont{F.}~\bibnamefont{Aubin}},
  \bibinfo{author}{\bibfnamefont{M.}~\bibnamefont{Auger}},
  \bibinfo{author}{\bibfnamefont{M.-H.} \bibnamefont{Genest}},
  \bibinfo{author}{\bibfnamefont{G.}~\bibnamefont{Giroux}},
  \bibinfo{author}{\bibfnamefont{R.}~\bibnamefont{Gornea}},
  \bibinfo{author}{\bibfnamefont{R.}~\bibnamefont{Faust}},
  \bibinfo{author}{\bibfnamefont{C.}~\bibnamefont{Leroy}},
  \bibinfo{author}{\bibfnamefont{L.}~\bibnamefont{Lessard}},
  \bibinfo{author}{\bibfnamefont{J.-P.} \bibnamefont{Martin}},
  \bibinfo{author}{\bibfnamefont{T.}~\bibnamefont{Morlat}},
  \bibnamefont{et~al.}, \bibinfo{journal}{New J. Phys.}
  \textbf{\bibinfo{volume}{10}}, \bibinfo{pages}{103017}
  (\bibinfo{year}{2008}).

\bibitem[{\citenamefont{Plesset and Zwick}(1954)}]{plesset_growth_1954}
\bibinfo{author}{\bibfnamefont{M.~S.} \bibnamefont{Plesset}} \bibnamefont{and}
  \bibinfo{author}{\bibfnamefont{S.~A.} \bibnamefont{Zwick}},
  \bibinfo{journal}{J. Appl. Phys.} \textbf{\bibinfo{volume}{25}},
  \bibinfo{pages}{493} (\bibinfo{year}{1954}).

\bibitem[{\citenamefont{Rayleigh}(1917)}]{rayleigh_pressure_1917}
\bibinfo{author}{\bibfnamefont{L.}~\bibnamefont{Rayleigh}},
  \bibinfo{journal}{Lond. Edinb. Dubl. Phil. Mag.}
  \textbf{\bibinfo{volume}{34}}, \bibinfo{pages}{94} (\bibinfo{year}{1917}).

\bibitem[{\citenamefont{Plimpton}(1995)}]{plimpton_fast_1995}
\bibinfo{author}{\bibfnamefont{S.}~\bibnamefont{Plimpton}},
  \bibinfo{journal}{J. Comput. Phys.} \textbf{\bibinfo{volume}{117}},
  \bibinfo{pages}{1} (\bibinfo{year}{1995}).

\bibitem[{\citenamefont{Denzel et~al.}(2016)\citenamefont{Denzel, Diemand, and
  Ang\'{e}lil}}]{denzel_molecular_2016}
\bibinfo{author}{\bibfnamefont{P.}~\bibnamefont{Denzel}},
  \bibinfo{author}{\bibfnamefont{J.}~\bibnamefont{Diemand}}, \bibnamefont{and}
  \bibinfo{author}{\bibfnamefont{R.}~\bibnamefont{Ang\'{e}lil}},
  \bibinfo{journal}{Phys. Rev. E} \textbf{\bibinfo{volume}{93}},
  \bibinfo{pages}{013301} (\bibinfo{year}{2016}).

\bibitem[{\citenamefont{Felizardo et~al.}(2012)\citenamefont{Felizardo, Girard,
  Morlat, Fernandes, Ramos, Marques, Kling, Puibasset, Auguste
  et~al.}}]{the_simple_collaboration_final_2012}
\bibinfo{author}{\bibfnamefont{M.}~\bibnamefont{Felizardo}},
  \bibinfo{author}{\bibfnamefont{T.~A.} \bibnamefont{Girard}},
  \bibinfo{author}{\bibfnamefont{T.}~\bibnamefont{Morlat}},
  \bibinfo{author}{\bibfnamefont{A.~C.} \bibnamefont{Fernandes}},
  \bibinfo{author}{\bibfnamefont{A.~R.} \bibnamefont{Ramos}},
  \bibinfo{author}{\bibfnamefont{J.~G.} \bibnamefont{Marques}},
  \bibinfo{author}{\bibfnamefont{A.}~\bibnamefont{Kling}},
  \bibinfo{author}{\bibfnamefont{J.}~\bibnamefont{Puibasset}},
  \bibinfo{author}{\bibfnamefont{M.}~\bibnamefont{Auguste}},
  \bibnamefont{et~al.} (\bibinfo{collaboration}{SIMPLE Collaboration}),
  \bibinfo{journal}{Phys. Rev. Lett.} \textbf{\bibinfo{volume}{108}},
  \bibinfo{pages}{201302} (\bibinfo{year}{2012}).

\bibitem[{\citenamefont{Lemmon et~al.}(2018)\citenamefont{Lemmon, Bell, Huber,
  and McLinden}}]{LEMMON-RP91}
\bibinfo{author}{\bibfnamefont{E.~W.} \bibnamefont{Lemmon}},
  \bibinfo{author}{\bibfnamefont{I.}~\bibnamefont{Bell}},
  \bibinfo{author}{\bibfnamefont{M.~L.} \bibnamefont{Huber}}, \bibnamefont{and}
  \bibinfo{author}{\bibfnamefont{M.~O.} \bibnamefont{McLinden}},
  \emph{\bibinfo{title}{{NIST Standard Reference Database 23: Reference Fluid
  Thermodynamic and Transport Properties-REFPROP, Version 9.0, National
  Institute of Standards and Technology}}} (\bibinfo{year}{2010}).

\bibitem[{\citenamefont{Seitz}(1958)}]{seitz_theory_1958}
\bibinfo{author}{\bibfnamefont{F.}~\bibnamefont{Seitz}},
  \bibinfo{journal}{Phys. Fluids} \textbf{\bibinfo{volume}{1}},
  \bibinfo{pages}{2} (\bibinfo{year}{1958}).

\bibitem[{\citenamefont{Lorensen et~al.}(1987)\citenamefont{Lorensen, Cline,
  Lorensen, and Cline}}]{lorensen_marching_1987}
\bibinfo{author}{\bibfnamefont{W.~E.} \bibnamefont{Lorensen}},
  \bibinfo{author}{\bibfnamefont{H.~E.} \bibnamefont{Cline}},
  \bibinfo{author}{\bibfnamefont{W.~E.} \bibnamefont{Lorensen}},
  \bibnamefont{and} \bibinfo{author}{\bibfnamefont{H.~E.} \bibnamefont{Cline}},
  \bibinfo{journal}{{ACM} {SIGGRAPH} Computer Graphics}
  \textbf{\bibinfo{volume}{21}}, \bibinfo{pages}{163} (\bibinfo{year}{1987}).

\bibitem[{\citenamefont{Walt et~al.}(2014)\citenamefont{Walt, Schönberger,
  Nunez-Iglesias, Boulogne, Warner, Yager, Gouillart, and
  Yu}}]{walt_scikit-image:_2014}
\bibinfo{author}{\bibfnamefont{S.~v.~d.} \bibnamefont{Walt}},
  \bibinfo{author}{\bibfnamefont{J.~L.} \bibnamefont{Schönberger}},
  \bibinfo{author}{\bibfnamefont{J.}~\bibnamefont{Nunez-Iglesias}},
  \bibinfo{author}{\bibfnamefont{F.}~\bibnamefont{Boulogne}},
  \bibinfo{author}{\bibfnamefont{J.~D.} \bibnamefont{Warner}},
  \bibinfo{author}{\bibfnamefont{N.}~\bibnamefont{Yager}},
  \bibinfo{author}{\bibfnamefont{E.}~\bibnamefont{Gouillart}},
  \bibnamefont{and} \bibinfo{author}{\bibfnamefont{T.}~\bibnamefont{Yu}},
  \bibinfo{journal}{{PeerJ}} \textbf{\bibinfo{volume}{2}},
  \bibinfo{pages}{e453} (\bibinfo{year}{2014}).

\bibitem[{\citenamefont{Landau and Lifshitz}(2013)}]{landau_fluid_2013}
\bibinfo{author}{\bibfnamefont{L.}~\bibnamefont{Landau}} \bibnamefont{and}
  \bibinfo{author}{\bibfnamefont{E.}~\bibnamefont{Lifshitz}},
  \emph{\bibinfo{title}{Fluid {Mechanics}}}, \bibinfo{number}{v. 6}
  (\bibinfo{publisher}{Elsevier Science}, \bibinfo{year}{2013}).

\bibitem[{\citenamefont{Mikic et~al.}(1970)\citenamefont{Mikic, Rohsenow, and
  Griffith}}]{mikic_bubble_1970}
\bibinfo{author}{\bibfnamefont{B.~B.} \bibnamefont{Mikic}},
  \bibinfo{author}{\bibfnamefont{W.~M.} \bibnamefont{Rohsenow}},
  \bibnamefont{and} \bibinfo{author}{\bibfnamefont{P.}~\bibnamefont{Griffith}},
  \bibinfo{journal}{Int. J. Heat Mass Transf.} \textbf{\bibinfo{volume}{13}},
  \bibinfo{pages}{657} (\bibinfo{year}{1970}).

\bibitem[{\citenamefont{Ziegler et~al.}(2010)\citenamefont{Ziegler, Ziegler,
  and Biersack}}]{ziegler_srim_2010}
\bibinfo{author}{\bibfnamefont{J.~F.} \bibnamefont{Ziegler}},
  \bibinfo{author}{\bibfnamefont{M.~D.} \bibnamefont{Ziegler}},
  \bibnamefont{and} \bibinfo{author}{\bibfnamefont{J.~P.}
  \bibnamefont{Biersack}}, \bibinfo{journal}{Nucl. Instrum. Meth. Phys. Res. B}
  \textbf{\bibinfo{volume}{268}}, \bibinfo{pages}{1818} (\bibinfo{year}{2010}).

\bibitem[{\citenamefont{Marsh et~al.}(1995)\citenamefont{Marsh, Thomas, and
  Burke}}]{marsh_high_1995}
\bibinfo{author}{\bibfnamefont{J.}~\bibnamefont{Marsh}},
  \bibinfo{author}{\bibfnamefont{D.}~\bibnamefont{Thomas}}, \bibnamefont{and}
  \bibinfo{author}{\bibfnamefont{M.}~\bibnamefont{Burke}},
  \bibinfo{journal}{Nucl. Instrum. Meth. Phys. Res. A}
  \textbf{\bibinfo{volume}{366}}, \bibinfo{pages}{340} (\bibinfo{year}{1995}).

\bibitem[{\citenamefont{Agostinelli et~al.}(2003)\citenamefont{Agostinelli,
  Allison, Amako, Apostolakis, Araujo, Arce, Asai, Axen, Banerjee, Barrand
  et~al.}}]{agostinelli_geant4simulation_2003}
\bibinfo{author}{\bibfnamefont{S.}~\bibnamefont{Agostinelli}},
  \bibinfo{author}{\bibfnamefont{J.}~\bibnamefont{Allison}},
  \bibinfo{author}{\bibfnamefont{K.}~\bibnamefont{Amako}},
  \bibinfo{author}{\bibfnamefont{J.}~\bibnamefont{Apostolakis}},
  \bibinfo{author}{\bibfnamefont{H.}~\bibnamefont{Araujo}},
  \bibinfo{author}{\bibfnamefont{P.}~\bibnamefont{Arce}},
  \bibinfo{author}{\bibfnamefont{M.}~\bibnamefont{Asai}},
  \bibinfo{author}{\bibfnamefont{D.}~\bibnamefont{Axen}},
  \bibinfo{author}{\bibfnamefont{S.}~\bibnamefont{Banerjee}},
  \bibinfo{author}{\bibfnamefont{G.}~\bibnamefont{Barrand}},
  \bibnamefont{et~al.}, \bibinfo{journal}{Nucl. Instrum. Meth. Phys. Res. A}
  \textbf{\bibinfo{volume}{506}}, \bibinfo{pages}{250} (\bibinfo{year}{2003}).

\bibitem[{\citenamefont{Brown et~al.}(2018)\citenamefont{Brown, Chadwick,
  Capote, Kahler, Trkov, Herman, Sonzogni, Danon, Carlson, Dunn
  et~al.}}]{brown_endf/b-viii.0:_2018}
\bibinfo{author}{\bibfnamefont{D.~A.} \bibnamefont{Brown}},
  \bibinfo{author}{\bibfnamefont{M.~B.} \bibnamefont{Chadwick}},
  \bibinfo{author}{\bibfnamefont{R.}~\bibnamefont{Capote}},
  \bibinfo{author}{\bibfnamefont{A.~C.} \bibnamefont{Kahler}},
  \bibinfo{author}{\bibfnamefont{A.}~\bibnamefont{Trkov}},
  \bibinfo{author}{\bibfnamefont{M.~W.} \bibnamefont{Herman}},
  \bibinfo{author}{\bibfnamefont{A.~A.} \bibnamefont{Sonzogni}},
  \bibinfo{author}{\bibfnamefont{Y.}~\bibnamefont{Danon}},
  \bibinfo{author}{\bibfnamefont{A.~D.} \bibnamefont{Carlson}},
  \bibinfo{author}{\bibfnamefont{M.}~\bibnamefont{Dunn}}, \bibnamefont{et~al.},
  \bibinfo{journal}{Nucl. Data Sheets} \textbf{\bibinfo{volume}{148}},
  \bibinfo{pages}{1} (\bibinfo{year}{2018}).

\bibitem[{\citenamefont{Amole
  et~al.}(2019{\natexlab{b}})}]{electron_recoil_paper}
\bibinfo{author}{\bibfnamefont{C.}~\bibnamefont{Amole}} \bibnamefont{et~al.}
  (\bibinfo{collaboration}{PICO Collaboration}),
  \bibinfo{journal}{arXiv:1905.12522 [physics.ins-det]}
  (\bibinfo{year}{2019}{\natexlab{b}}).

\bibitem[{\citenamefont{Jones}(1924)}]{jones_determination_1924}
\bibinfo{author}{\bibfnamefont{J.~E.} \bibnamefont{Jones}},
  \bibinfo{journal}{Proc. R. Soc. Lond. A} \textbf{\bibinfo{volume}{106}},
  \bibinfo{pages}{463} (\bibinfo{year}{1924}).

\bibitem[{\citenamefont{Toxvaerd and
  Dyre}(2011)}]{toxvaerd_communication:_2011}
\bibinfo{author}{\bibfnamefont{S.}~\bibnamefont{Toxvaerd}} \bibnamefont{and}
  \bibinfo{author}{\bibfnamefont{J.~C.} \bibnamefont{Dyre}},
  \bibinfo{journal}{J. Chem. Phys.} \textbf{\bibinfo{volume}{134}},
  \bibinfo{pages}{081102} (\bibinfo{year}{2011}).

\bibitem[{\citenamefont{Diemand et~al.}(2014)\citenamefont{Diemand,
  Ang\'{e}lil, Tanaka, and Tanaka}}]{diemand_direct_2014}
\bibinfo{author}{\bibfnamefont{J.}~\bibnamefont{Diemand}},
  \bibinfo{author}{\bibfnamefont{R.}~\bibnamefont{Ang\'{e}lil}},
  \bibinfo{author}{\bibfnamefont{K.~K.} \bibnamefont{Tanaka}},
  \bibnamefont{and} \bibinfo{author}{\bibfnamefont{H.}~\bibnamefont{Tanaka}},
  \bibinfo{journal}{Phys. Rev. E} \textbf{\bibinfo{volume}{90}},
  \bibinfo{pages}{052407} (\bibinfo{year}{2014}).

\bibitem[{\citenamefont{Walton et~al.}(1983)\citenamefont{Walton, Tildesley,
  Rowlinson, and Henderson}}]{walton_pressure_1983}
\bibinfo{author}{\bibfnamefont{J.~P. R.~B.} \bibnamefont{Walton}},
  \bibinfo{author}{\bibfnamefont{D.~J.} \bibnamefont{Tildesley}},
  \bibinfo{author}{\bibfnamefont{J.~S.} \bibnamefont{Rowlinson}},
  \bibnamefont{and} \bibinfo{author}{\bibfnamefont{J.~R.}
  \bibnamefont{Henderson}}, \bibinfo{journal}{Mol. Phys.}
  \textbf{\bibinfo{volume}{48}}, \bibinfo{pages}{1357} (\bibinfo{year}{1983}).

\bibitem[{\citenamefont{Chapela et~al.}(1977)\citenamefont{Chapela, Saville,
  Thompson, and Rowlinson}}]{chapela_computer_1977}
\bibinfo{author}{\bibfnamefont{G.~A.} \bibnamefont{Chapela}},
  \bibinfo{author}{\bibfnamefont{G.}~\bibnamefont{Saville}},
  \bibinfo{author}{\bibfnamefont{S.~M.} \bibnamefont{Thompson}},
  \bibnamefont{and} \bibinfo{author}{\bibfnamefont{J.~S.}
  \bibnamefont{Rowlinson}}, \bibinfo{journal}{J. Chem. Soc. Faraday Trans. 2}
  \textbf{\bibinfo{volume}{73}}, \bibinfo{pages}{1133} (\bibinfo{year}{1977}).

\bibitem[{\citenamefont{Kirkwood and Buff}(1949)}]{kirkwood_statistical_1949}
\bibinfo{author}{\bibfnamefont{J.~G.} \bibnamefont{Kirkwood}} \bibnamefont{and}
  \bibinfo{author}{\bibfnamefont{F.~P.} \bibnamefont{Buff}},
  \bibinfo{journal}{J. Chem. Phys.} \textbf{\bibinfo{volume}{17}},
  \bibinfo{pages}{338} (\bibinfo{year}{1949}).

\end{thebibliography}

 \newcommand{\noop}[1]{}

\end{document}